\documentclass[longauth]{aaEC}
\usepackage{comment}
\usepackage{graphicx}
\usepackage{natbib}
\bibpunct{(}{)}{,}{a}{}{;}
\usepackage{scalerel}
\usepackage{multirow}
\usepackage{booktabs,tabularx}
\usepackage[table]{xcolor}
\usepackage{newtxtext,newtxmath}
\usepackage{amsfonts,amsmath,amssymb,mathtools,bm,multirow}
\usepackage{adjustbox}
\usepackage[normalem]{ulem}
\usepackage{cancel}
\usepackage{upgreek} 

\usepackage{siunitx}
\DeclareSIUnit\jansky{Jy}

\usepackage[pdfencoding=auto,psdextra]{hyperref}
\hypersetup{
    colorlinks=true,
    linkcolor=blue,
    filecolor=magenta,      
    urlcolor=blue,
    citecolor=blue
}
\usepackage[nameinlink,capitalise]{cleveref}
\crefname{section}{Sect.}{Sects}
\Crefname{section}{Section}{Sections}
\makeatletter
\renewcommand*\aa@pageof{, page \thepage{} of \pageref*{LastPage}}
\makeatother
\definecolor{ForestGreen}{HTML}{009B55}
\definecolor{Gold}{HTML}{B59410}

\newcommand{\de}{\mathrm{d}}

\newcommand{\snr}{\ensuremath{\text{S/N}}}
\newcommand{\selavy}{\texttt{Selavy}}
\newcommand{\pybdsf}{\texttt{PyBDSF}}
\newcommand{\pgal}{\texttt{pgal0.90}}
\newcommand{\pstar}{\texttt{pstar0.05}}

\let\Gamma\varGamma
\let\Delta\varDelta
\let\Theta\varTheta
\let\Xi\varXi
\let\Pi\varPi
\let\Upsilon\varUpsilon
\let\Phi\varPhi
\let\Psi\varPsi
\let\Omega\varOmega

\usepackage[utf8]{inputenc}

\usepackage[switch, modulo]{lineno}

\usepackage{euclid}

\begin{document}

%
%

\title{EMU and \Euclid: Detection of a radio-optical galaxy clustering cross-correlation signal between the Evolutionary Map of the Universe and \Euclid\thanks{This paper is published on behalf of the EMU Collaboration and the Euclid Consortium.}}


\newcommand{\orcid}[1]{} 
\author{G.~Piccirilli\orcid{0000-0002-3341-1872}\thanks{\email{giulia.piccirilli@roma2.infn.it}}\inst{\ref{aff1},\ref{aff2}}
\and B.~Bahr-Kalus\orcid{0000-0002-4578-4019}\inst{\ref{aff3},\ref{aff1},\ref{aff2}}
\and S.~Camera\orcid{0000-0003-3399-3574}\inst{\ref{aff1},\ref{aff2},\ref{aff3}}
\and J.~Asorey\orcid{0000-0002-6211-499X}\inst{\ref{aff4}}
\and C.~L.~Hale\inst{\ref{aff5},\ref{aff6}}
\and G.~Fabbian\orcid{0000-0002-3255-4695}\inst{\ref{aff7}}
\and A.~D.~Asher\orcid{0000-0003-0854-7732}\inst{\ref{aff8},\ref{aff9}}
\and M.~Vai\inst{\ref{aff1},\ref{aff2}}
\and C.~S.~Saraf\orcid{0000-0002-5149-4042}\inst{\ref{aff10}}
\and D.~Parkinson\orcid{0000-0002-7464-2351}\inst{\ref{aff10},\ref{aff11}}
\and N.~Tessore\orcid{0000-0002-9696-7931}\inst{\ref{aff12},\ref{aff13}}
\and K.~Tanidis\orcid{0000-0001-9843-5130}\inst{\ref{aff5},\ref{aff14}}
\and M.~Kunz\orcid{0000-0002-3052-7394}\inst{\ref{aff15}}
\and A.~M.~Hopkins\orcid{0000-0002-6097-2747}\inst{\ref{aff16}}
\and T.~Vernstrom\orcid{0000-0001-7093-3875}\inst{\ref{aff17},\ref{aff18}}
\and M.~Regis\orcid{0000-0003-0399-0284}\inst{\ref{aff1},\ref{aff2}}
\and M.~J.~I.~Brown\orcid{0000-0002-1207-9137}\inst{\ref{aff19}}
\and D.~Carollo\orcid{0000-0002-0005-5787}\inst{\ref{aff20}}
\and T.~Zafar\orcid{0000-0003-3935-7018}\inst{\ref{aff16}}
\and R.~P.~Norris\orcid{0000-0002-4597-1906}\inst{\ref{aff9},\ref{aff8}}
\and F.~Pace\orcid{0000-0001-8039-0480}\inst{\ref{aff1},\ref{aff2},\ref{aff3}}
\and J.~M.~Diego\orcid{0000-0001-9065-3926}\inst{\ref{aff21}}
\and H.~Tang\orcid{0000-0002-7300-9239}\inst{\ref{aff22}}
\and F.~Rahman\orcid{0000-0001-9414-175X}\inst{\ref{aff23}}
\and D.~Farrah\orcid{0000-0003-1748-2010}\inst{\ref{aff24},\ref{aff25}}
\and J.~Th.~van~Loon\orcid{0000-0002-1272-3017}\inst{\ref{aff26}}
\and C.~M.~Pennock\orcid{0000-0003-2164-5337}\inst{\ref{aff6}}
\and J.~Willingham\orcid{0009-0001-5653-9481}\inst{\ref{aff16},\ref{aff27}}
\and S.~Andreon\inst{\ref{aff28}}
\and C.~Baccigalupi\orcid{0000-0002-8211-1630}\inst{\ref{aff29},\ref{aff20},\ref{aff30},\ref{aff31}}
\and M.~Baldi\orcid{0000-0003-4145-1943}\inst{\ref{aff32},\ref{aff33},\ref{aff34}}
\and S.~Bardelli\orcid{0000-0002-8900-0298}\inst{\ref{aff33}}
\and A.~Biviano\orcid{0000-0002-0857-0732}\inst{\ref{aff20},\ref{aff29}}
\and E.~Branchini\orcid{0000-0002-0808-6908}\inst{\ref{aff35},\ref{aff36},\ref{aff28}}
\and M.~Brescia\orcid{0000-0001-9506-5680}\inst{\ref{aff37},\ref{aff38}}
\and G.~Ca\~nas-Herrera\orcid{0000-0003-2796-2149}\inst{\ref{aff39},\ref{aff40}}
\and V.~Capobianco\orcid{0000-0002-3309-7692}\inst{\ref{aff3}}
\and C.~Carbone\orcid{0000-0003-0125-3563}\inst{\ref{aff41}}
\and V.~F.~Cardone\inst{\ref{aff42},\ref{aff43}}
\and J.~Carretero\orcid{0000-0002-3130-0204}\inst{\ref{aff44},\ref{aff45}}
\and S.~Casas\orcid{0000-0002-4751-5138}\inst{\ref{aff46},\ref{aff47}}
\and M.~Castellano\orcid{0000-0001-9875-8263}\inst{\ref{aff42}}
\and G.~Castignani\orcid{0000-0001-6831-0687}\inst{\ref{aff33}}
\and S.~Cavuoti\orcid{0000-0002-3787-4196}\inst{\ref{aff38},\ref{aff48}}
\and K.~C.~Chambers\orcid{0000-0001-6965-7789}\inst{\ref{aff25}}
\and A.~Cimatti\inst{\ref{aff49}}
\and C.~Colodro-Conde\inst{\ref{aff50}}
\and G.~Congedo\orcid{0000-0003-2508-0046}\inst{\ref{aff6}}
\and L.~Conversi\orcid{0000-0002-6710-8476}\inst{\ref{aff51},\ref{aff52}}
\and Y.~Copin\orcid{0000-0002-5317-7518}\inst{\ref{aff53}}
\and F.~Courbin\orcid{0000-0003-0758-6510}\inst{\ref{aff54},\ref{aff55},\ref{aff56}}
\and H.~M.~Courtois\orcid{0000-0003-0509-1776}\inst{\ref{aff57}}
\and M.~Cropper\orcid{0000-0003-4571-9468}\inst{\ref{aff12}}
\and A.~Da~Silva\orcid{0000-0002-6385-1609}\inst{\ref{aff58},\ref{aff59}}
\and H.~Degaudenzi\orcid{0000-0002-5887-6799}\inst{\ref{aff60}}
\and G.~De~Lucia\orcid{0000-0002-6220-9104}\inst{\ref{aff20}}
\and H.~Dole\orcid{0000-0002-9767-3839}\inst{\ref{aff7}}
\and M.~Douspis\orcid{0000-0003-4203-3954}\inst{\ref{aff7}}
\and F.~Dubath\orcid{0000-0002-6533-2810}\inst{\ref{aff60}}
\and C.~A.~J.~Duncan\orcid{0009-0003-3573-0791}\inst{\ref{aff6}}
\and X.~Dupac\inst{\ref{aff52}}
\and S.~Dusini\orcid{0000-0002-1128-0664}\inst{\ref{aff61}}
\and S.~Escoffier\orcid{0000-0002-2847-7498}\inst{\ref{aff62}}
\and M.~Farina\orcid{0000-0002-3089-7846}\inst{\ref{aff63}}
\and R.~Farinelli\inst{\ref{aff33}}
\and F.~Faustini\orcid{0000-0001-6274-5145}\inst{\ref{aff42},\ref{aff64}}
\and S.~Ferriol\inst{\ref{aff53}}
\and F.~Finelli\orcid{0000-0002-6694-3269}\inst{\ref{aff33},\ref{aff65}}
\and M.~Frailis\orcid{0000-0002-7400-2135}\inst{\ref{aff20}}
\and E.~Franceschi\orcid{0000-0002-0585-6591}\inst{\ref{aff33}}
\and M.~Fumana\orcid{0000-0001-6787-5950}\inst{\ref{aff41}}
\and S.~Galeotta\orcid{0000-0002-3748-5115}\inst{\ref{aff20}}
\and K.~George\orcid{0000-0002-1734-8455}\inst{\ref{aff66}}
\and B.~Gillis\orcid{0000-0002-4478-1270}\inst{\ref{aff6}}
\and C.~Giocoli\orcid{0000-0002-9590-7961}\inst{\ref{aff33},\ref{aff34}}
\and J.~Gracia-Carpio\inst{\ref{aff67}}
\and A.~Grazian\orcid{0000-0002-5688-0663}\inst{\ref{aff68}}
\and F.~Grupp\inst{\ref{aff67},\ref{aff69}}
\and L.~Guzzo\orcid{0000-0001-8264-5192}\inst{\ref{aff70},\ref{aff28},\ref{aff71}}
\and S.~V.~H.~Haugan\orcid{0000-0001-9648-7260}\inst{\ref{aff72}}
\and W.~Holmes\inst{\ref{aff73}}
\and I.~M.~Hook\orcid{0000-0002-2960-978X}\inst{\ref{aff74}}
\and F.~Hormuth\inst{\ref{aff75}}
\and A.~Hornstrup\orcid{0000-0002-3363-0936}\inst{\ref{aff76},\ref{aff77}}
\and K.~Jahnke\orcid{0000-0003-3804-2137}\inst{\ref{aff78}}
\and M.~Jhabvala\inst{\ref{aff79}}
\and B.~Joachimi\orcid{0000-0001-7494-1303}\inst{\ref{aff13}}
\and E.~Keih\"anen\orcid{0000-0003-1804-7715}\inst{\ref{aff80}}
\and S.~Kermiche\orcid{0000-0002-0302-5735}\inst{\ref{aff62}}
\and A.~Kiessling\orcid{0000-0002-2590-1273}\inst{\ref{aff73}}
\and M.~Kilbinger\orcid{0000-0001-9513-7138}\inst{\ref{aff81}}
\and B.~Kubik\orcid{0009-0006-5823-4880}\inst{\ref{aff53}}
\and M.~K\"ummel\orcid{0000-0003-2791-2117}\inst{\ref{aff69}}
\and H.~Kurki-Suonio\orcid{0000-0002-4618-3063}\inst{\ref{aff82},\ref{aff83}}
\and A.~M.~C.~Le~Brun\orcid{0000-0002-0936-4594}\inst{\ref{aff84}}
\and S.~Ligori\orcid{0000-0003-4172-4606}\inst{\ref{aff3}}
\and P.~B.~Lilje\orcid{0000-0003-4324-7794}\inst{\ref{aff72}}
\and V.~Lindholm\orcid{0000-0003-2317-5471}\inst{\ref{aff82},\ref{aff83}}
\and I.~Lloro\orcid{0000-0001-5966-1434}\inst{\ref{aff85}}
\and G.~Mainetti\orcid{0000-0003-2384-2377}\inst{\ref{aff86}}
\and D.~Maino\inst{\ref{aff70},\ref{aff41},\ref{aff71}}
\and O.~Mansutti\orcid{0000-0001-5758-4658}\inst{\ref{aff20}}
\and S.~Marcin\inst{\ref{aff87}}
\and O.~Marggraf\orcid{0000-0001-7242-3852}\inst{\ref{aff88}}
\and M.~Martinelli\orcid{0000-0002-6943-7732}\inst{\ref{aff42},\ref{aff43}}
\and N.~Martinet\orcid{0000-0003-2786-7790}\inst{\ref{aff89}}
\and F.~Marulli\orcid{0000-0002-8850-0303}\inst{\ref{aff90},\ref{aff33},\ref{aff34}}
\and R.~J.~Massey\orcid{0000-0002-6085-3780}\inst{\ref{aff91}}
\and E.~Medinaceli\orcid{0000-0002-4040-7783}\inst{\ref{aff33}}
\and S.~Mei\orcid{0000-0002-2849-559X}\inst{\ref{aff92},\ref{aff93}}
\and Y.~Mellier\inst{\ref{aff94},\ref{aff95}}
\and M.~Meneghetti\orcid{0000-0003-1225-7084}\inst{\ref{aff33},\ref{aff34}}
\and E.~Merlin\orcid{0000-0001-6870-8900}\inst{\ref{aff42}}
\and G.~Meylan\inst{\ref{aff96}}
\and A.~Mora\orcid{0000-0002-1922-8529}\inst{\ref{aff97}}
\and M.~Moresco\orcid{0000-0002-7616-7136}\inst{\ref{aff90},\ref{aff33}}
\and L.~Moscardini\orcid{0000-0002-3473-6716}\inst{\ref{aff90},\ref{aff33},\ref{aff34}}
\and R.~Nakajima\orcid{0009-0009-1213-7040}\inst{\ref{aff88}}
\and C.~Neissner\orcid{0000-0001-8524-4968}\inst{\ref{aff98},\ref{aff45}}
\and R.~C.~Nichol\orcid{0000-0003-0939-6518}\inst{\ref{aff99}}
\and S.-M.~Niemi\orcid{0009-0005-0247-0086}\inst{\ref{aff39}}
\and C.~Padilla\orcid{0000-0001-7951-0166}\inst{\ref{aff98}}
\and K.~Paech\orcid{0000-0003-0625-2367}\inst{\ref{aff67}}
\and S.~Paltani\orcid{0000-0002-8108-9179}\inst{\ref{aff60}}
\and F.~Pasian\orcid{0000-0002-4869-3227}\inst{\ref{aff20}}
\and K.~Pedersen\inst{\ref{aff100}}
\and W.~J.~Percival\orcid{0000-0002-0644-5727}\inst{\ref{aff101},\ref{aff102},\ref{aff103}}
\and V.~Pettorino\orcid{0000-0002-4203-9320}\inst{\ref{aff39}}
\and S.~Pires\orcid{0000-0002-0249-2104}\inst{\ref{aff81}}
\and G.~Polenta\orcid{0000-0003-4067-9196}\inst{\ref{aff64}}
\and M.~Poncet\inst{\ref{aff104}}
\and L.~A.~Popa\inst{\ref{aff105}}
\and L.~Pozzetti\orcid{0000-0001-7085-0412}\inst{\ref{aff33}}
\and F.~Raison\orcid{0000-0002-7819-6918}\inst{\ref{aff67}}
\and A.~Renzi\orcid{0000-0001-9856-1970}\inst{\ref{aff106},\ref{aff61}}
\and J.~Rhodes\orcid{0000-0002-4485-8549}\inst{\ref{aff73}}
\and G.~Riccio\inst{\ref{aff38}}
\and E.~Romelli\orcid{0000-0003-3069-9222}\inst{\ref{aff20}}
\and M.~Roncarelli\orcid{0000-0001-9587-7822}\inst{\ref{aff33}}
\and R.~Saglia\orcid{0000-0003-0378-7032}\inst{\ref{aff69},\ref{aff67}}
\and D.~Sapone\orcid{0000-0001-7089-4503}\inst{\ref{aff107}}
\and B.~Sartoris\orcid{0000-0003-1337-5269}\inst{\ref{aff69},\ref{aff20}}
\and J.~A.~Schewtschenko\orcid{0000-0002-4913-6393}\inst{\ref{aff6}}
\and P.~Schneider\orcid{0000-0001-8561-2679}\inst{\ref{aff88}}
\and T.~Schrabback\orcid{0000-0002-6987-7834}\inst{\ref{aff108}}
\and A.~Secroun\orcid{0000-0003-0505-3710}\inst{\ref{aff62}}
\and G.~Seidel\orcid{0000-0003-2907-353X}\inst{\ref{aff78}}
\and S.~Serrano\orcid{0000-0002-0211-2861}\inst{\ref{aff109},\ref{aff110},\ref{aff111}}
\and P.~Simon\inst{\ref{aff88}}
\and C.~Sirignano\orcid{0000-0002-0995-7146}\inst{\ref{aff106},\ref{aff61}}
\and G.~Sirri\orcid{0000-0003-2626-2853}\inst{\ref{aff34}}
\and A.~Spurio~Mancini\orcid{0000-0001-5698-0990}\inst{\ref{aff112}}
\and L.~Stanco\orcid{0000-0002-9706-5104}\inst{\ref{aff61}}
\and J.-L.~Starck\orcid{0000-0003-2177-7794}\inst{\ref{aff81}}
\and J.~Steinwagner\orcid{0000-0001-7443-1047}\inst{\ref{aff67}}
\and P.~Tallada-Cresp\'{i}\orcid{0000-0002-1336-8328}\inst{\ref{aff44},\ref{aff45}}
\and A.~N.~Taylor\inst{\ref{aff6}}
\and I.~Tereno\orcid{0000-0002-4537-6218}\inst{\ref{aff58},\ref{aff113}}
\and S.~Toft\orcid{0000-0003-3631-7176}\inst{\ref{aff114},\ref{aff115}}
\and R.~Toledo-Moreo\orcid{0000-0002-2997-4859}\inst{\ref{aff116}}
\and F.~Torradeflot\orcid{0000-0003-1160-1517}\inst{\ref{aff45},\ref{aff44}}
\and I.~Tutusaus\orcid{0000-0002-3199-0399}\inst{\ref{aff111},\ref{aff109},\ref{aff117}}
\and L.~Valenziano\orcid{0000-0002-1170-0104}\inst{\ref{aff33},\ref{aff65}}
\and J.~Valiviita\orcid{0000-0001-6225-3693}\inst{\ref{aff82},\ref{aff83}}
\and T.~Vassallo\orcid{0000-0001-6512-6358}\inst{\ref{aff20}}
\and A.~Veropalumbo\orcid{0000-0003-2387-1194}\inst{\ref{aff28},\ref{aff36},\ref{aff35}}
\and Y.~Wang\orcid{0000-0002-4749-2984}\inst{\ref{aff118}}
\and J.~Weller\orcid{0000-0002-8282-2010}\inst{\ref{aff69},\ref{aff67}}
\and G.~Zamorani\orcid{0000-0002-2318-301X}\inst{\ref{aff33}}
\and F.~M.~Zerbi\inst{\ref{aff28}}
\and E.~Zucca\orcid{0000-0002-5845-8132}\inst{\ref{aff33}}
\and J.~Garc\'ia-Bellido\orcid{0000-0002-9370-8360}\inst{\ref{aff119}}
\and J.~Mart\'{i}n-Fleitas\orcid{0000-0002-8594-569X}\inst{\ref{aff120}}
\and A.~Pezzotta\orcid{0000-0003-0726-2268}\inst{\ref{aff28}}
\and V.~Scottez\orcid{0009-0008-3864-940X}\inst{\ref{aff94},\ref{aff121}}
\and M.~Viel\orcid{0000-0002-2642-5707}\inst{\ref{aff29},\ref{aff20},\ref{aff31},\ref{aff30},\ref{aff122}}}
										   
\institute{Dipartimento di Fisica, Universit\`a degli Studi di Torino, Via P. Giuria 1, 10125 Torino, Italy\label{aff1}
\and
INFN-Sezione di Torino, Via P. Giuria 1, 10125 Torino, Italy\label{aff2}
\and
INAF-Osservatorio Astrofisico di Torino, Via Osservatorio 20, 10025 Pino Torinese (TO), Italy\label{aff3}
\and
Departamento de F\'{\i}sica Te\'orica, Centro de Astropart\'iculas y F\'isica de Altas Energ\'ias (CAPA), Universidad de Zaragoza, 50009 Zaragoza, Spain\label{aff4}
\and
Department of Physics, Oxford University, Keble Road, Oxford OX1 3RH, UK\label{aff5}
\and
Institute for Astronomy, University of Edinburgh, Royal Observatory, Blackford Hill, Edinburgh EH9 3HJ, UK\label{aff6}
\and
Universit\'e Paris-Saclay, CNRS, Institut d'astrophysique spatiale, 91405, Orsay, France\label{aff7}
\and
Western Sydney University, Locked Bag 1797, Penrith South DC, NSW 2751, Australia\label{aff8}
\and
Australia Telescope National Facility, CSIRO, Space and Astronomy, PO Box 76, Epping, NSW 1710, Australia\label{aff9}
\and
Korea Astronomy and Space Science Institute, 776 Daedeok-daero, Yuseong-gu, Daejeon 34055, Republic of Korea\label{aff10}
\and
University of Science and Technology, 217 Gajeong-ro, Yuseong-gu, Daejeon 34113, Republic of Korea\label{aff11}
\and
Mullard Space Science Laboratory, University College London, Holmbury St Mary, Dorking, Surrey RH5 6NT, UK\label{aff12}
\and
Department of Physics and Astronomy, University College London, Gower Street, London WC1E 6BT, UK\label{aff13}
\and
Center for Astrophysics and Cosmology, University of Nova Gorica, Nova Gorica, Slovenia\label{aff14}
\and
Universit\'e de Gen\`eve, D\'epartement de Physique Th\'eorique and Centre for Astroparticle Physics, 24 quai Ernest-Ansermet, CH-1211 Gen\`eve 4, Switzerland\label{aff15}
\and
School of Mathematical and Physical Sciences, Macquarie University, Sydney, NSW 2109, Australia\label{aff16}
\and
ATNF, CSIRO, Space and Astronomy, PO Box 1130, Bentley, WA 6151, Australia\label{aff17}
\and
International Centre for Radio Astronomy Research, University of Western Australia, 35 Stirling Highway, Crawley, Western Australia 6009, Australia\label{aff18}
\and
School of Physics and Astronomy, Monash University, Clayton, VIC 3800, Australia\label{aff19}
\and
INAF-Osservatorio Astronomico di Trieste, Via G. B. Tiepolo 11, 34143 Trieste, Italy\label{aff20}
\and
Instituto de F\'isica de Cantabria, Edificio Juan Jord\'a, Avenida de los Castros, 39005 Santander, Spain\label{aff21}
\and
Department of Physics, Xi'an Jiaotong-Liverpool University, Suzhou 215123, China\label{aff22}
\and
Syed Babar Ali School of Science and Engineering (SBASSE), Lahore University of Management Sciences (LUMS), Lahore, Pakistan\label{aff23}
\and
Department of Physics and Astronomy, University of Hawai`i at Manoa, 2505 Correa Rd., Honolulu, HI, 96822, USA\label{aff24}
\and
Institute for Astronomy, University of Hawaii, 2680 Woodlawn Drive, Honolulu, HI 96822, USA\label{aff25}
\and
Lennard-Jones Laboratories, Keele University, ST5 5BG, UK\label{aff26}
\and
Macquarie University Astrophysics and Space Technologies Research Centre, Sydney, NSW 2109, Australia\label{aff27}
\and
INAF-Osservatorio Astronomico di Brera, Via Brera 28, 20122 Milano, Italy\label{aff28}
\and
IFPU, Institute for Fundamental Physics of the Universe, via Beirut 2, 34151 Trieste, Italy\label{aff29}
\and
INFN, Sezione di Trieste, Via Valerio 2, 34127 Trieste TS, Italy\label{aff30}
\and
SISSA, International School for Advanced Studies, Via Bonomea 265, 34136 Trieste TS, Italy\label{aff31}
\and
Dipartimento di Fisica e Astronomia, Universit\`a di Bologna, Via Gobetti 93/2, 40129 Bologna, Italy\label{aff32}
\and
INAF-Osservatorio di Astrofisica e Scienza dello Spazio di Bologna, Via Piero Gobetti 93/3, 40129 Bologna, Italy\label{aff33}
\and
INFN-Sezione di Bologna, Viale Berti Pichat 6/2, 40127 Bologna, Italy\label{aff34}
\and
Dipartimento di Fisica, Universit\`a di Genova, Via Dodecaneso 33, 16146, Genova, Italy\label{aff35}
\and
INFN-Sezione di Genova, Via Dodecaneso 33, 16146, Genova, Italy\label{aff36}
\and
Department of Physics "E. Pancini", University Federico II, Via Cinthia 6, 80126, Napoli, Italy\label{aff37}
\and
INAF-Osservatorio Astronomico di Capodimonte, Via Moiariello 16, 80131 Napoli, Italy\label{aff38}
\and
European Space Agency/ESTEC, Keplerlaan 1, 2201 AZ Noordwijk, The Netherlands\label{aff39}
\and
Leiden Observatory, Leiden University, Einsteinweg 55, 2333 CC Leiden, The Netherlands\label{aff40}
\and
INAF-IASF Milano, Via Alfonso Corti 12, 20133 Milano, Italy\label{aff41}
\and
INAF-Osservatorio Astronomico di Roma, Via Frascati 33, 00078 Monteporzio Catone, Italy\label{aff42}
\and
INFN-Sezione di Roma, Piazzale Aldo Moro, 2 - c/o Dipartimento di Fisica, Edificio G. Marconi, 00185 Roma, Italy\label{aff43}
\and
Centro de Investigaciones Energ\'eticas, Medioambientales y Tecnol\'ogicas (CIEMAT), Avenida Complutense 40, 28040 Madrid, Spain\label{aff44}
\and
Port d'Informaci\'{o} Cient\'{i}fica, Campus UAB, C. Albareda s/n, 08193 Bellaterra (Barcelona), Spain\label{aff45}
\and
Institute for Theoretical Particle Physics and Cosmology (TTK), RWTH Aachen University, 52056 Aachen, Germany\label{aff46}
\and
Deutsches Zentrum f\"ur Luft- und Raumfahrt e. V. (DLR), Linder H\"ohe, 51147 K\"oln, Germany\label{aff47}
\and
INFN section of Naples, Via Cinthia 6, 80126, Napoli, Italy\label{aff48}
\and
Dipartimento di Fisica e Astronomia "Augusto Righi" - Alma Mater Studiorum Universit\`a di Bologna, Viale Berti Pichat 6/2, 40127 Bologna, Italy\label{aff49}
\and
Instituto de Astrof\'{\i}sica de Canarias, E-38205 La Laguna, Tenerife, Spain\label{aff50}
\and
European Space Agency/ESRIN, Largo Galileo Galilei 1, 00044 Frascati, Roma, Italy\label{aff51}
\and
ESAC/ESA, Camino Bajo del Castillo, s/n., Urb. Villafranca del Castillo, 28692 Villanueva de la Ca\~nada, Madrid, Spain\label{aff52}
\and
Universit\'e Claude Bernard Lyon 1, CNRS/IN2P3, IP2I Lyon, UMR 5822, Villeurbanne, F-69100, France\label{aff53}
\and
Institut de Ci\`{e}ncies del Cosmos (ICCUB), Universitat de Barcelona (IEEC-UB), Mart\'{i} i Franqu\`{e}s 1, 08028 Barcelona, Spain\label{aff54}
\and
Instituci\'o Catalana de Recerca i Estudis Avan\c{c}ats (ICREA), Passeig de Llu\'{\i}s Companys 23, 08010 Barcelona, Spain\label{aff55}
\and
Institut de Ciencies de l'Espai (IEEC-CSIC), Campus UAB, Carrer de Can Magrans, s/n Cerdanyola del Vall\'es, 08193 Barcelona, Spain\label{aff56}
\and
UCB Lyon 1, CNRS/IN2P3, IUF, IP2I Lyon, 4 rue Enrico Fermi, 69622 Villeurbanne, France\label{aff57}
\and
Departamento de F\'isica, Faculdade de Ci\^encias, Universidade de Lisboa, Edif\'icio C8, Campo Grande, PT1749-016 Lisboa, Portugal\label{aff58}
\and
Instituto de Astrof\'isica e Ci\^encias do Espa\c{c}o, Faculdade de Ci\^encias, Universidade de Lisboa, Campo Grande, 1749-016 Lisboa, Portugal\label{aff59}
\and
Department of Astronomy, University of Geneva, ch. d'Ecogia 16, 1290 Versoix, Switzerland\label{aff60}
\and
INFN-Padova, Via Marzolo 8, 35131 Padova, Italy\label{aff61}
\and
Aix-Marseille Universit\'e, CNRS/IN2P3, CPPM, Marseille, France\label{aff62}
\and
INAF-Istituto di Astrofisica e Planetologia Spaziali, via del Fosso del Cavaliere, 100, 00100 Roma, Italy\label{aff63}
\and
Space Science Data Center, Italian Space Agency, via del Politecnico snc, 00133 Roma, Italy\label{aff64}
\and
INFN-Bologna, Via Irnerio 46, 40126 Bologna, Italy\label{aff65}
\and
University Observatory, LMU Faculty of Physics, Scheinerstr.~1, 81679 Munich, Germany\label{aff66}
\and
Max Planck Institute for Extraterrestrial Physics, Giessenbachstr. 1, 85748 Garching, Germany\label{aff67}
\and
INAF-Osservatorio Astronomico di Padova, Via dell'Osservatorio 5, 35122 Padova, Italy\label{aff68}
\and
Universit\"ats-Sternwarte M\"unchen, Fakult\"at f\"ur Physik, Ludwig-Maximilians-Universit\"at M\"unchen, Scheinerstr.~1, 81679 M\"unchen, Germany\label{aff69}
\and
Dipartimento di Fisica "Aldo Pontremoli", Universit\`a degli Studi di Milano, Via Celoria 16, 20133 Milano, Italy\label{aff70}
\and
INFN-Sezione di Milano, Via Celoria 16, 20133 Milano, Italy\label{aff71}
\and
Institute of Theoretical Astrophysics, University of Oslo, P.O. Box 1029 Blindern, 0315 Oslo, Norway\label{aff72}
\and
Jet Propulsion Laboratory, California Institute of Technology, 4800 Oak Grove Drive, Pasadena, CA, 91109, USA\label{aff73}
\and
Department of Physics, Lancaster University, Lancaster, LA1 4YB, UK\label{aff74}
\and
Felix Hormuth Engineering, Goethestr. 17, 69181 Leimen, Germany\label{aff75}
\and
Technical University of Denmark, Elektrovej 327, 2800 Kgs. Lyngby, Denmark\label{aff76}
\and
Cosmic Dawn Center (DAWN), Denmark\label{aff77}
\and
Max-Planck-Institut f\"ur Astronomie, K\"onigstuhl 17, 69117 Heidelberg, Germany\label{aff78}
\and
NASA Goddard Space Flight Center, Greenbelt, MD 20771, USA\label{aff79}
\and
Department of Physics and Helsinki Institute of Physics, Gustaf H\"allstr\"omin katu 2, University of Helsinki, 00014 Helsinki, Finland\label{aff80}
\and
Universit\'e Paris-Saclay, Universit\'e Paris Cit\'e, CEA, CNRS, AIM, 91191, Gif-sur-Yvette, France\label{aff81}
\and
Department of Physics, P.O. Box 64, University of Helsinki, 00014 Helsinki, Finland\label{aff82}
\and
Helsinki Institute of Physics, Gustaf H{\"a}llstr{\"o}min katu 2, University of Helsinki, 00014 Helsinki, Finland\label{aff83}
\and
Laboratoire d'etude de l'Univers et des phenomenes eXtremes, Observatoire de Paris, Universit\'e PSL, Sorbonne Universit\'e, CNRS, 92190 Meudon, France\label{aff84}
\and
SKAO, Jodrell Bank, Lower Withington, Macclesfield SK11 9FT, UK\label{aff85}
\and
Centre de Calcul de l'IN2P3/CNRS, 21 avenue Pierre de Coubertin 69627 Villeurbanne Cedex, France\label{aff86}
\and
University of Applied Sciences and Arts of Northwestern Switzerland, School of Computer Science, 5210 Windisch, Switzerland\label{aff87}
\and
Universit\"at Bonn, Argelander-Institut f\"ur Astronomie, Auf dem H\"ugel 71, 53121 Bonn, Germany\label{aff88}
\and
Aix-Marseille Universit\'e, CNRS, CNES, LAM, Marseille, France\label{aff89}
\and
Dipartimento di Fisica e Astronomia "Augusto Righi" - Alma Mater Studiorum Universit\`a di Bologna, via Piero Gobetti 93/2, 40129 Bologna, Italy\label{aff90}
\and
Department of Physics, Institute for Computational Cosmology, Durham University, South Road, Durham, DH1 3LE, UK\label{aff91}
\and
Universit\'e Paris Cit\'e, CNRS, Astroparticule et Cosmologie, 75013 Paris, France\label{aff92}
\and
CNRS-UCB International Research Laboratory, Centre Pierre Bin\'etruy, IRL2007, CPB-IN2P3, Berkeley, USA\label{aff93}
\and
Institut d'Astrophysique de Paris, 98bis Boulevard Arago, 75014, Paris, France\label{aff94}
\and
Institut d'Astrophysique de Paris, UMR 7095, CNRS, and Sorbonne Universit\'e, 98 bis boulevard Arago, 75014 Paris, France\label{aff95}
\and
Institute of Physics, Laboratory of Astrophysics, Ecole Polytechnique F\'ed\'erale de Lausanne (EPFL), Observatoire de Sauverny, 1290 Versoix, Switzerland\label{aff96}
\and
Telespazio UK S.L. for European Space Agency (ESA), Camino bajo del Castillo, s/n, Urbanizacion Villafranca del Castillo, Villanueva de la Ca\~nada, 28692 Madrid, Spain\label{aff97}
\and
Institut de F\'{i}sica d'Altes Energies (IFAE), The Barcelona Institute of Science and Technology, Campus UAB, 08193 Bellaterra (Barcelona), Spain\label{aff98}
\and
School of Mathematics and Physics, University of Surrey, Guildford, Surrey, GU2 7XH, UK\label{aff99}
\and
DARK, Niels Bohr Institute, University of Copenhagen, Jagtvej 155, 2200 Copenhagen, Denmark\label{aff100}
\and
Waterloo Centre for Astrophysics, University of Waterloo, Waterloo, Ontario N2L 3G1, Canada\label{aff101}
\and
Department of Physics and Astronomy, University of Waterloo, Waterloo, Ontario N2L 3G1, Canada\label{aff102}
\and
Perimeter Institute for Theoretical Physics, Waterloo, Ontario N2L 2Y5, Canada\label{aff103}
\and
Centre National d'Etudes Spatiales -- Centre spatial de Toulouse, 18 avenue Edouard Belin, 31401 Toulouse Cedex 9, France\label{aff104}
\and
Institute of Space Science, Str. Atomistilor, nr. 409 M\u{a}gurele, Ilfov, 077125, Romania\label{aff105}
\and
Dipartimento di Fisica e Astronomia "G. Galilei", Universit\`a di Padova, Via Marzolo 8, 35131 Padova, Italy\label{aff106}
\and
Departamento de F\'isica, FCFM, Universidad de Chile, Blanco Encalada 2008, Santiago, Chile\label{aff107}
\and
Universit\"at Innsbruck, Institut f\"ur Astro- und Teilchenphysik, Technikerstr. 25/8, 6020 Innsbruck, Austria\label{aff108}
\and
Institut d'Estudis Espacials de Catalunya (IEEC),  Edifici RDIT, Campus UPC, 08860 Castelldefels, Barcelona, Spain\label{aff109}
\and
Satlantis, University Science Park, Sede Bld 48940, Leioa-Bilbao, Spain\label{aff110}
\and
Institute of Space Sciences (ICE, CSIC), Campus UAB, Carrer de Can Magrans, s/n, 08193 Barcelona, Spain\label{aff111}
\and
Department of Physics, Royal Holloway, University of London, Surrey TW20 0EX, UK\label{aff112}
\and
Instituto de Astrof\'isica e Ci\^encias do Espa\c{c}o, Faculdade de Ci\^encias, Universidade de Lisboa, Tapada da Ajuda, 1349-018 Lisboa, Portugal\label{aff113}
\and
Cosmic Dawn Center (DAWN)\label{aff114}
\and
Niels Bohr Institute, University of Copenhagen, Jagtvej 128, 2200 Copenhagen, Denmark\label{aff115}
\and
Universidad Polit\'ecnica de Cartagena, Departamento de Electr\'onica y Tecnolog\'ia de Computadoras,  Plaza del Hospital 1, 30202 Cartagena, Spain\label{aff116}
\and
Institut de Recherche en Astrophysique et Plan\'etologie (IRAP), Universit\'e de Toulouse, CNRS, UPS, CNES, 14 Av. Edouard Belin, 31400 Toulouse, France\label{aff117}
\and
Infrared Processing and Analysis Center, California Institute of Technology, Pasadena, CA 91125, USA\label{aff118}
\and
Instituto de F\'isica Te\'orica UAM-CSIC, Campus de Cantoblanco, 28049 Madrid, Spain\label{aff119}
\and
Aurora Technology for European Space Agency (ESA), Camino bajo del Castillo, s/n, Urbanizacion Villafranca del Castillo, Villanueva de la Ca\~nada, 28692 Madrid, Spain\label{aff120}
\and
ICL, Junia, Universit\'e Catholique de Lille, LITL, 59000 Lille, France\label{aff121}
\and
ICSC - Centro Nazionale di Ricerca in High Performance Computing, Big Data e Quantum Computing, Via Magnanelli 2, Bologna, Italy\label{aff122}}    

%
\abstract{
Synergies between large-scale radio-continuum and optical/near-infrared galaxy surveys have long been recognised as a powerful tool for cosmology. Cross-correlating these surveys can constrain the redshift distribution of radio sources, mitigate systematic effects, and place strong constraints on cosmological models.
We perform the first measurement of the clustering cross-spectrum between radio-continuum sources in the Evolutionary Map of the Universe (EMU) survey and galaxies from the ESA \Euclid satellite mission's Q1 release. Our goal is to detect and characterise the cross-correlation signal, test its robustness against systematic effects, and compare our measurements with theoretical predictions.
We use data from the Australian SKA Pathfinder's (ASKAP) EMU Main Survey, which overlaps with the \(28\,\deg^2\)  Euclid Deep Field South. We generate two radio-source catalogues using different source finders to create galaxy maps. We measure the harmonic-space cross-correlation signal using a pseudo-spectrum estimator for angular multipoles up to \(\ell\lesssim800\), roughly corresponding to angular separations above \(13 \farcm 5\). The measured signal is compared to theoretical predictions based on a fiducial \(\Lambda\)CDM cosmology, using several models for the EMU source redshift distribution and bias.
We report detection above \(8\,\sigma\) of the cross-correlation signal consistent across all tested models and data sets. The measured cross-spectra from the two independently generated radio catalogues are in excellent agreement, demonstrating that the cross-correlation is robust against the choice of source-finding algorithm, a key potential systematic error. The measured signal also agrees with theoretical models developed from previous cross-correlation studies and simulations at the \(68\%\) confidence level.
This pathfinder study establishes a statistically significant cross-correlation between EMU and \Euclid. The robustness of the signal is a crucial validation of the methodology, paving the way for future large-scale analyses leveraging the full power of this synergy to constrain cosmological parameters and our understanding of galaxy evolution.
    }
%
%
%
%

\titlerunning{First detection of EMU-\Euclid cross-correlation}
\authorrunning{Piccirilli et al.}
   
\nolinenumbers
\maketitle
%
%
%
%
\section{\label{sc:Intro}Introduction}
The concordance cosmological model, \(\Lambda\)CDM, has been remarkably successful in describing the evolution and large-scale properties of the Universe. However, the physical nature of its two main components, dark matter and dark energy, remains one of the most profound mysteries in modern physics, amplified by recent results favouring the time evolution of the latter \citep{DESI:dr1baocosmo, DESI:dr2baocosmo}.

To test the \(\Lambda\)CDM paradigm and measure precisely its parameters, we can rely on mapping the cosmic large-scale structure (LSS). The spatial distribution of matter, as traced by galaxies and other astrophysical objects, is a powerful repository of information about the initial conditions of the Universe, its expansion history, and the growth of cosmic structures over time. Dark matter governs the rate at which matter clusters under gravity, determining the amplitude of density fluctuations over cosmic time. Meanwhilst, dark energy influences the expansion rate of the Universe, which acts as a brake on the growth of these structures. Therefore, mapping the evolution of the galaxy density field provides a powerful test of the \(\Lambda\)CDM paradigm and allows us to measure fundamental cosmological parameters like the matter abundance, \(\Omega_{\rm m}\), and the dark energy equation of state.

Upcoming and ongoing large-volume surveys are set to map the LSS with unprecedented precision, opening a new era of observational cosmology. A particularly powerful technique in this endeavour is the multi-tracer approach \citep[e.g.][]{Seljak:2009,2009JCAP...10..007M,2013MNRAS.432..318A,2014MNRAS.442.2511F,2015PhRvD..92f3525A,Fonseca:multitracer}, which involves combining data from different surveys that trace the same underlying matter density field. By cross-correlating different populations of objects (e.g., galaxies observed at different wavelengths), one can mitigate the effects of cosmic variance, lift the degeneracy between astrophysical bias and fundamental cosmological parameters, and gain tighter control over survey-specific systematic errors. This is particularly crucial for studies of ultra-large-scale effects such as the scale-dependent bias effect or the integrated Sachs--Wolfe effect, whose measurements suffer from cosmic variance limitations and are prone to large-scale systematics \citep[e.g.][]{2022MNRAS.517.3785B,2024MNRAS.532.1902R}.

A promising synergy is between radio and optical/near-infrared galaxy surveys \citep[e.g.][]{2012MNRAS.427.2079C,2023MNRAS.518.6262C,2025A&A...698A..58Z}. With that in mind, this work focuses on two landmark cosmological surveys: ESA's \Euclid satellite mission and the Evolutionary Map of the Universe (EMU). \Euclid \citep{Laureijs11, EuclidSkyOverview} is designed to map the geometry of the dark Universe by measuring the shapes and redshifts of billions of galaxies out to \(z\approx 2.5\) and beyond, over a third of the sky. Its primary goals are to constrain the nature of dark energy and test general relativity
by using weak gravitational lensing and galaxy clustering. In addition to its wide survey, \Euclid repeatedly observes several Deep Fields to provide a deeper, multi-faceted view of the cosmos.

Complementing \Euclid's optical and near-infrared view, the EMU survey \citep{2011PASA...28..215N, Hopkins:EMUmain} is creating a deep, high-resolution radio atlas of the entire Southern Sky,\footnote{The final survey footprint will not be exactly, but roughly, delimited by the equator.} using the Australian SKA Pathfinder (ASKAP) telescope. EMU is expected to detect approximately 20 million radio sources, predominantly star-forming galaxies (SFGs) and active galactic nuclei (AGN). As a radio continuum survey, EMU provides excellent angular positions but does not measure the redshift for its sources. This presents a significant challenge for using EMU for three-dimensional cosmological analyses.

The cross-correlation between \Euclid and EMU offers a compelling solution to this challenge and a powerful probe of cosmology in its own right. Since both surveys trace the same underlying LSS, their cross-spectrum is expected to have a high signal-to-noise ratio (\snr). This signal can be used to statistically constrain the redshift distribution of the EMU radio sources, a technique known as clustering redshifts \citep[e.g.][]{Menard:clusteringredshifts}. By dividing the \Euclid sample into redshift bins, one can perform a tomographic analysis of the cross-correlation to reconstruct the redshift distribution and galaxy bias of EMU galaxies, and even their halo properties \citep{Krolewski:TomoHOD}. This, in turn, unlocks the potential of the EMU survey for cosmological studies. Furthermore, as AGN-dominated radio galaxy samples are typically more clustered than the general galaxy population \citep[e.g.][]{2017MNRAS.464.3271M,Hale2018,2019A&A...623A.148D}, their clustering signal provides a potent lever arm for cosmological measurements, especially when using the multi-tracer approach.

In this paper, we present the first measurement of the harmonic-space cross-spectrum between the EMU Main Survey and \Euclid. We focus on the region of the \Euclid Deep Field South, which is included in \Euclid's first `quick release' (Q1)---a non-cosmology release of Level 2 data meant for astrophysical studies. We measure the clustering signal between EMU and \Euclid galaxies and compare it with theoretical models based on different assumptions for the radio source redshift distribution and galaxy bias, building upon previous work that cross-correlated radio-continuum surveys, such as EMU, with Stage-III optical surveys.\footnote{For instance, VISTA Deep Extragalactic Observations cross-correlated with the Canada–France–Hawaii Telescope Legacy Survey \citep{Lindsay_2014}, EMU cross-correlated with the Dark Energy Survey (DES, \citealt{Saraf:EMUxDES}), and the Low Frequency Array (LOFAR) Two-Metre Sky Survey (LoTSS) cross-correlated with the extended Baryon Oscillation Spectroscopic Survey (eBOSS, \citealt{2025A&A...698A..58Z}).}

A key aspect of our analysis is to test the robustness of the cross-correlation signal against potential systematics in radio data processing, specifically by comparing results derived from two independent source-finding algorithms. Additionally, we use the cross-correlation signal to measure \Euclid's galaxy bias, confirming the findings from our companion paper by Fabbian et al.\ (in prep.), which measured the same quantity from correlations of \Euclid with CMB data. 

This work serves as a pathfinder for future, larger-scale analyses combining the full \Euclid and EMU data sets, demonstrating the scientific potential of this powerful synergy. Whilst we want to use \Euclid and EMU to constrain cosmology once larger datasets are available, here, we assume a flat \(\Lambda\)CDM fiducial cosmology consistent with the Planck 2018 results \citep{2020A&A...641A...6P} throughout this paper, with parameters: \(H_0=\qty{67.4}{\kilo\meter\per\second\per\mega\parsec}\), \(\Omega_\mathrm{m}=0.315\), and \(\Omega_{\rm b}=0.049\).

The paper is structured as follows. In \cref{sc:data_surveys}, we describe the \Euclid and EMU data sets used in this analysis. \Cref{ssec:theory_met.theory} outlines the theoretical framework for our modelling and \cref{sec:methodology} the power spectrum estimation methodology. We present our measurements of the harmonic-space power spectra and the derived constraints in \cref{sec:results}. We discuss our findings and conclude in \cref{sec:conclusions}.

\section{Data}
\label{sc:data_surveys}
Here, we describe the data sets employed in this analysis.
We process each data set in a similar way. We start by generating the projected galaxy density contrast \(\Delta\) on a map using the Hierarchical Equal Area isoLatitude Pixelation of a sphere \citep[\texttt{HEALPix; }][]{2005ApJ...622..759G,Zonca2019},\footnote{\url{https://healpix.sourceforge.io/}} with resolution parameter \(N_{\rm side} = 1024\) (corresponding to approximately \(\ang{;3.42}\)), chosen as the highest resolution where we have on average more than one EMU object per pixel. We then define the projected density contrast as
\begin{equation}
\label{eq:delta_g}
    \Delta(\hat{\vec n}) = \frac{N_w(\hat{\vec n})}{\bar N_w} -1\;,
\end{equation}
where \(\hat{\vec n}\) is the direction of the \(p\)th pixel, \(N_w(\hat{\vec n}) = N(\hat{\vec n})/w(\hat{\vec n})\) is the weighted number of sources, with \(N(\hat{\vec n})\) the raw observed number of sources in that pixel, and \(w(\hat{\vec n})\) the value of the selection function of the galaxy sample in that same pixel---thus, also accounting for the overall mask. Then, \(\bar N_w\) is the mean weighted number of galaxies per pixel, computed as
\begin{equation}\label{eq:Nbar}
\bar N_w = \frac{1}{N_\mathrm{pix}}\sum_p N_w(\hat{\vec n})\;,
\end{equation}
where \(N_\mathrm{pix}\) is the number of pixels in the survey footprint.
The \Euclid weights are further described in Fabbian et al.\ (in prep.) whilst we describe how we obtain weights for EMU in \cref{sec:EMUdata}. 

\subsection{The \Euclid Q1 Catalogue}\label{sec:Eucliddata}
The \Euclid data used in this work are from the first Quick Data Release \citep[][Q1 hereafter]{2025arXiv250315302E}. We focus on the Euclid Deep Field South (EDF-S), a \(28\,\deg^2\) field centred around \(({\rm RA}, {\rm Dec}) = (61\overset{\circ}{.}241, -48\overset{\circ}{.}423)\). This initial release is primarily intended for astrophysical studies and sample characterisation. Our analysis relies on the photometric source catalogue generated from a combination of \Euclid's VIS \citep{EuclidSkyVIS} and NISP instruments \citep{EuclidSkyNISP}, supplemented by extensive ground-based ancillary data in the \(g\), \(r\), \(i\), and \(z\) bands, which together span a total wavelength coverage of approximately \(0.55\)--\(2.0\,\si{\micro\meter}\) and provide the robust photometric redshift estimates required for this work.

For our cross-correlation analysis, the redshift information for the \Euclid sources is paramount. In particular, we use the same photometric redshifts (photo-\(z\)s) and classification as the ones in Fabbian et al.\ (in prep.).
Indeed, together with the redshift information, the Q1 catalogue includes object classification in terms of probabilities of an object being a galaxy (\(p_{\rm gal}\)), a QSO (\(p_{\rm qso}\)), or a star (\(p_{\rm star}\)).\footnote{Note that radio continuum emission from stars is rare in Stokes I, as shown by LoTSS \citep{2022A&A...659A...1S,2023A&A...670A.124C} and the Rapid ASKAP Continuum Survey \citep[RACS;][]{2021MNRAS.502.5438P,2023PASA...40...36D}. This should be confirmed with EMU's sister project, the Polarisation Sky Survey of the Universe's Magnetism (POSSUM), but for now, we assume that the EMU signal is only from galaxies and that star contamination in \Euclid will drop out in cross-correlation. It might, however, affect the overall amplitude of the cross-correlation, since it will modify the normalisation of the overdensity map.\label{fn:stars}} In particular, we consider two main samples selected by the probability cuts \(p_{\rm gal} > 0.90\) (\pgal, hereafter) and \(p_{\rm star} < 0.05\) (\pstar). The final galaxy number densities for these cuts are approximately \(11\,{\rm arcmin}^{-2}\) and \(21\,{\rm arcmin}^{-2}\), respectively. Following the selection, we combine all qualifying galaxies into a single, wide redshift bin, whilst in \cref{sec:Nzfit}, we will make use of three tomographic redshift bins with the same redshift range as Fabbian et al.\ (in prep.), i.e., with edges at redshifts \(0,0.5,0.8\), and \(2.5\).

Finally, we create a galaxy overdensity map from the positions of the selected galaxies using \cref{eq:delta_g} and a binary selection function, i.e. \(w(\hat{\vec n}) = 1\) for observed pixels and \(w(\hat{\vec n}) = 0\), everywhere else, with some apodisation. As in Fabbian et al.\ (in prep.), the observational systematics are not included in a single selection function but systematic modes are deprojected before estimating the angular power spectra. 
As the deprojection effectively alters the total number of observed objects (see e.g. \citealt{2019MNRAS.482..453K}), we renormalise the deprojected density contrast \(\Delta^\prime(\hat{\vec n})\) using 
\begin{equation}
\label{eq:delta_g_v2}
    \Delta(\hat{\vec n}) = \frac{1 + \Delta^\prime(\hat{\vec n})}{\left\langle 1 + \Delta^\prime(\hat{\vec n})\right\rangle} - 1
\end{equation}
to always ensure a zero-mean distributed \(\Delta(\hat{\vec n})\).\footnote{We verified that the measured \Euclid-EMU cross angular power spectrum is not significantly affected by the different definitions of \(\Delta(\hat{\vec n})\) in \cref{eq:delta_g,eq:delta_g_v2}.}
This serves as the input for the power spectrum estimation described in \cref{sec:Clestimation}.

\subsection{The EMU Main Survey}\label{sec:EMUdata}

The EMU project \citep{2011PASA...28..215N,2021PASA...38...46N, Hopkins:EMUmain}\footnote{\url{https://emu-survey.org/}.} uses the ASKAP radio telescope \citep{2021PASA...38....9H} to generate a deep radio atlas of the Southern sky, expecting to catalogue about 20 million extragalactic radio objects by 2028. We use publicly available data from four scheduling blocks (SBs), SB51431, SB59442, and SB71454, that overlap with EDF-S.\footnote{The data can be retrieved from the Commonwealth Scientific and Industrial Research Organisation's (CSIRO) ASKAP Science Data Archive (CASDA): \url{https://research.csiro.au/casda/}.} The observations were taken at a central frequency of \(\SI{943}{\mega\hertz}\) with a \(\SI{288}{\mega\hertz}\) bandwidth and restored to a common \(\ang{;;15}\) resolution. We exclude SB50787 due to minimal overlap and significant image artefacts.

A key systematic in high-resolution radio surveys is the presence of multi-component sources: a single physical galaxy (e.g., a core with jets and lobes) can be resolved by ASKAP and catalogued as multiple distinct entries. The use of such components for clustering analysis artificially boosts the small-scale power, contaminating the cosmological signal. To ensure each unique galaxy is counted only once, we employ island-level or source-level catalogues from two independent source finders:
\begin{itemize}
\item \selavy\ \citep{Whiting2012}: We use the public island catalogue, where each island groups the multiple Gaussian components associated with a single contiguous region of emission.
\item \pybdsf\ \citep{2015ascl.soft02007M}:\footnote{\url{https://pybdsf.readthedocs.io/en/latest/index.html}.} We generate an alternative source catalogue using standard detection thresholds (\(3\,\sigma\) for islands, \(5\,\sigma\) for sources) and robustly estimating the local noise (r.m.s.\ map). As source-finder uncertainty is a known systematic \citep{2025PASA...42...62T}, comparing results from both \selavy\ and \pybdsf\ is essential for validating our cross-spectrum measurement. A more in-depth comparison of source finders is currently being prepared by Barnes et al.\ (in prep.).\end{itemize}

Since a fully processed `supermosaic' image covering the EDF-S is not yet available,\footnote{As the EMU survey progresses, supermosaics will be produced in regions of enough adjacent SBs and will be available through the EMU Radio and Value-added Catalogue \citep[EMUCAT; ][]{Marvil2023} to EMU Team Members.} we combine the individual source catalogues from the three overlapping scheduling blocks. To mitigate potential double-counting in the overlap regions, we include sources within a given \texttt{HEALPix} cell only from the SB catalogue that exhibits the highest selection function value \(w(\hat{\vec n})\) (defined in the following paragraph) in that pixel. To define our sample, we apply a flux density cut of \(\qty{140}{\micro\jansky}\), which is five times the median r.m.s.\ noise per source. To ensure reliable flux measurement (critical for this cut), we leverage the component-level information: for both \selavy\ and \pybdsf, the total flux density of a unique source is calculated as the sum of the integrated flux densities of all its associated Gaussian components. This model-based approach is more robust against noise fluctuations than simple pixel summation.

\begin{figure}
    \centering
    \includegraphics[width=\columnwidth]{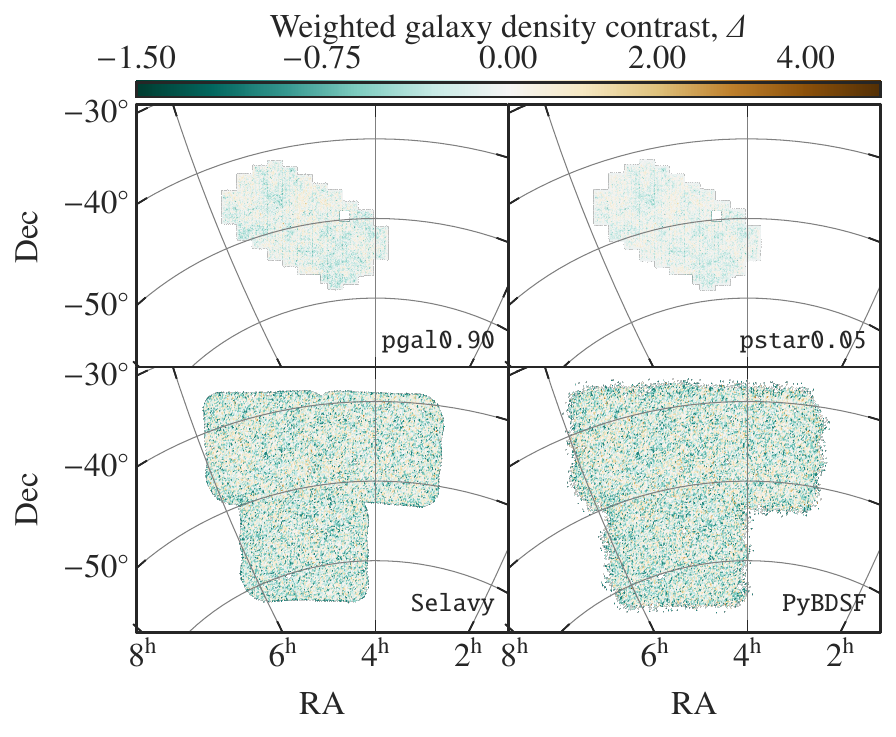}
    \caption{Weighted galaxy overdensity maps for \Euclid (\emph{top panels}: left for \pgal\ sample and right from \pstar\ sample) and EMU (\emph{bottom panels}: left for the \selavy\ source finder and right for the \pybdsf\ source finder) obtained with \cref{eq:delta_g_v2}.}
    \label{fig:footprints}
\end{figure}

The EMU survey effective sensitivity is not spatially uniform. To correct for variations in completeness, we construct a data weighting function, \(w(\hat{\vec n})\), for each \texttt{HEALPix} pixel. Following \citet{Hale2021,2022MNRAS.517.3785B,2025PASA...42...62T}, and \citet{Saraf:EMUxDES}, this weight is derived using source injection simulations based on the SKADS model \citep{2008MNRAS.388.1335W, 2010MNRAS.405..447W}, modified to increase the number of SFGs, where \(w(\hat{\vec n})\) is the ratio of simulated sources recovered above a \(5\,\sigma\) threshold to the number of injected sources in that pixel. We exclude all pixels with very low completeness, set to \(w(\hat{\vec n}) < 0.5\), from the final analysis. The final overdensity maps are shown in \cref{fig:footprints}. Whilst this simulation-based weighting has known limitations (e.g., in precisely capturing source finder performance), we argue that this is consistent with previous analyses and that for the current cross-spectrum measurement, the impact on the signal scale-dependence is subdominant, as the method depends on the relative variations between the two surveys. However, we acknowledge these limitations do influence the total signal normalisation and the signal-to-noise ratio by altering the effective number of sources, a factor implicitly absorbed into the fitted amplitude \(A_b\). We plan to use refined supermosaics and simulations in future analyses to provide a more robust treatment for the more sensitive EMU auto-spectrum measurement.

\section{Theoretical modelling}

\label{ssec:theory_met.theory}
In this analysis, we considered two projected quantities: the angular distribution of the \Euclid galaxies and the angular distribution of EMU radio sources. In real space, both radio sources and galaxies are biased tracers of the same underlying matter distribution. In a flat universe, their angular number counts fluctuations, \(\Delta(\hat{\vec n})\), is thus related to the comoving matter density contrast, \(\delta(\vec r)\), via
\begin{equation}
    \Delta(\hat{\vec n}) = \int \de z \, b(z) \, p(z) \, \delta[r(z)\,\hat{\vec n}]\;,\label{eq:Delta}
\end{equation}
where \(b(z)\) is the galaxy bias, \(r(z)\) is the radial comoving distance, such that \(\de r/\de z=c/H(z)\), \(H(z)\) being the expansion Hubble rate and \(c\) the speed of light, and \(p(z)=n(z)/\bar n\), with \(n(z)\) being the angular redshift distribution of the sources and \(\bar n\) their angular number density. In other words, \(p(z)\,\de z\) is the probability of finding a galaxy in the redshift interval \([z,z+\de z]\).

In principle, one should account for other contributions to \(\Delta\) besides matter density perturbations, because galaxy observations are inevitably performed not in real space but on the past light-cone. Such additional terms include the well-known redshift-space distortions \citep[RSD;][]{1987MNRAS.227....1K,1996ApJ...462...25Z} and lensing magnification \citep{1967ApJ...150..737G,1989Sci...245..824B,1997MNRAS.291..446D}, as well as others, usually dubbed `relativistic' effects \citep{Yoo:2010ni,Bonvin:2011bg,Challinor:2011bk}. However, the latter are largely subdominant, and important only on the largest cosmic scales, whereas our patch is very small. The contribution of RSD, instead, albeit important, is effectively washed out in harmonic-space power spectra with broad redshift windows \citep[see e.g.][]{2019MNRAS.489.3385T,2020MNRAS.491.4869T,Tanidis-TBD}. 

Thus, the harmonic-space power spectrum of galaxy clustering---i.e., the two-point correlation function of the harmonic coefficients of maps of fluctuations in galaxy number counts---in our case reads
\begin{equation}
    C_\ell^{AB} \simeq \int\frac{c\,\de z}{H(z)}\,W_A(z)\,W_B(z)\,P_{\rm m}\left(k = \frac{\ell+1/2}{r(z)}, z\right)\;,\label{eq:Cl_Limber}
\end{equation}
where, given the broad redshift bins, we used the so-called Limber approximation \citep{1953ApJ...117..134L,1998ApJ...498...26K}.
In the previous equation, \(P_{\rm m}\) is the matter power spectrum and \(W_A\) is the window function for tracer \(A\), namely
\begin{equation}\label{eq:kernel}
    W_A(z) = \frac{H(z)\,b_A(z)\,p_A(z)}{c\,r(z)}\;.
\end{equation}
Therefore, for each survey, we need to model their bias and redshift distributions. We note that, for both of them we will consider a linear bias since the scales we are considering are well within the quasi-linear regime (cf. \cref{sec:lmax_nonlin}). 

For \Euclid, the galaxy redshift distribution \(n(z)\) for each of the two samples (\texttt{pgal0.90} and \texttt{pstar0.05}, cf. \cref{sec:Eucliddata} and Fabbian et al., in prep.) is constructed by stacking the individual photo-\(z\) probability distribution functions (PDFs) of all galaxies in the final sample.\footnote{We acknowledge that this estimate might be biased \citep{2020arXiv200712178M}, but we find this bias to be negligible compared to our current uncertainty in the EMU redshift distribution.} These PDFs are derived from a full Bayesian hierarchical model fitting, as implemented in the Euclid Q1 pipeline \citep{Q1-TP005}. Whilst point redshift estimates may be used to define redshift bins (using the \texttt{PHZ\_MODE\_1} photo-\(z\) that represents the primary peak of the redshift probability distribution for each galaxy, i.e. the most likely redshift value), the \(n(z)\) itself is obtained by aggregating these full PDFs to accurately account for redshift uncertainties. The linear galaxy bias model is taken from the Flagship-2 simulation \citep{EuclidSkyFlagship}.
\begin{figure}
    \centering
    \includegraphics[width=\columnwidth]{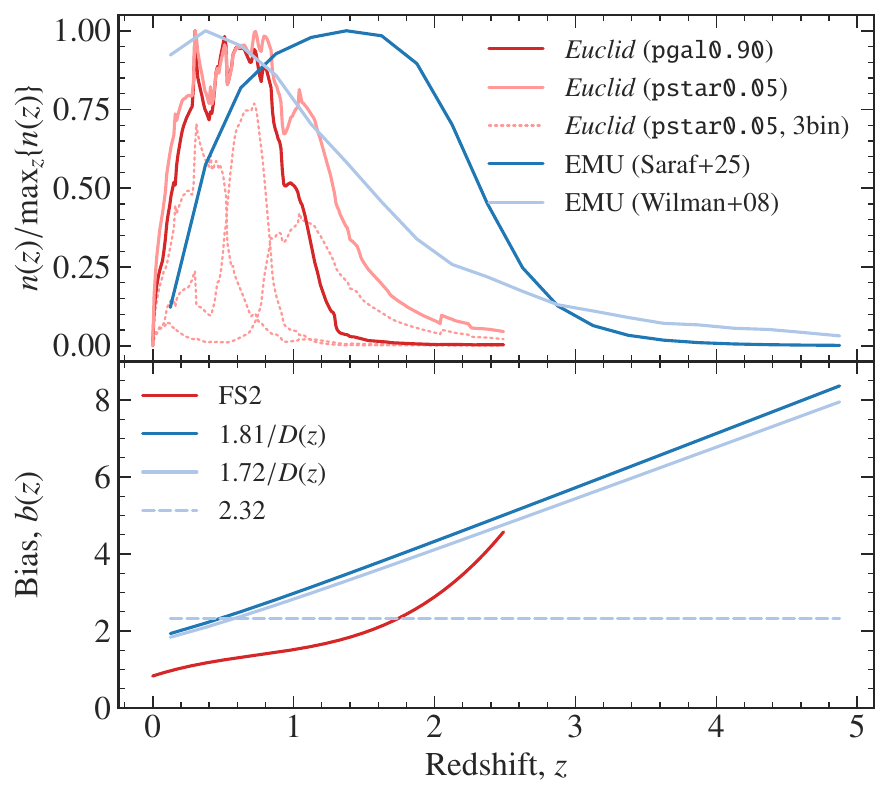}
    \caption{{\it Top panel}: Galaxy redshift distributions used in this analysis. For \Euclid (red curves), we show the {\tt 1bin} and {\tt 3bin} configurations with solid and dotted lines, respectively. For  EMU (blue curves) we plot two different models from \cite{Saraf:EMUxDES} and \cite{2008MNRAS.388.1335W}. Note that the unbinned redshift distributions are normalised to their peak, with the binned ones being normalised to the peak of the corresponding unbinned distribution, to enhance readability. {\it Bottom panel}: Fiducial biases adopted, for \Euclid galaxies from Flagship-2 (red curve) and for EMU sources considering both the constant bias model (blue, dashed line) and constant-amplitude one (blue, solid curves). 
    }
    \label{fig:dndz_bias}\label{fig:tomography}
\end{figure}

For EMU, instead, we consider different models for both the redshift source distribution and the bias. In particular, we use results from \citet{Saraf:EMUxDES} who performed a tomographic analysis of the cross-correlation between the EMU Pilot Survey 1 \citep{2021PASA...38...46N} and DES magnitude-limited galaxy sample \citep[\texttt{MagLim},][]{2021PhRvD.103d3503P, 2022PhRvD.106j3530P}. They provide an analytical formula for the EMU galaxy redshift distribution and a bias fit to \(1.81/D(z)\), where \(D(z)\) is the linear growth factor normalised to unity today. 
Besides, we also compare the data to the model and fit(s) in \citet{2025PASA...42...62T}, who assumed the \(n(z)\) from the SKADS simulations \citep{2008MNRAS.388.1335W} combined with two bias assumptions: a constant model, best-fit to \(2.32\), and a constant-amplitude model, with best-fit \(1.72/D(z)\). 
The value of these bias parameters are obtained from the combined analysis of the auto-spectrum of galaxies in the EMU Pilot Survey 1 and their cross-correlation with CMB lensing from \Planck. All the quantities described above are shown in \cref{fig:dndz_bias}, where a significant difference between the \(n_{\rm EMU}(z)\) models of \citet{2008MNRAS.388.1335W} and \citet{Saraf:EMUxDES} can be observed. We shall use the cross-correlation signal between EMU and \Euclid\ in \cref{sec:Nzfit} to obtain better understanding on what the true \(n_{\rm EMU}(z)\) should be.
To summarise, in our analysis we shall always use (and later compare and relax by fitting for a free amplitude parameter), for \Euclid, \(n(z)\) from Q1 data and FS2 bias, and for EMU, the following models:\footnote{We note that the analyses of \cite{2025PASA...42...62T} and \cite{Saraf:EMUxDES} adopt a slightly shallower flux density cut, \(S > 0.18\,\rm{mJy}\). We verified that the resulting cross-angular power spectra of \Euclid galaxies with EMU sources at that flux density cut are in agreement well within \(1\,\sigma\) error bars with our baseline analysis.}
\begin{itemize}
    \item {\it Model 1}: analytical \(n_{\rm EMU}(z)\) and \(b_{\rm EMU}(z)= 1.81/D(z)\), both from \cite{Saraf:EMUxDES};
    \item {\it Model 2}: SKADS \(n_{\rm EMU}(z)\) from \citet{2008MNRAS.388.1335W} and \(b_{\rm EMU}(z) = 1.72/D(z)\) from \citet{2025PASA...42...62T};
    \item {\it Model 3}: SKADS \(n_{\rm EMU}(z)\) from \citet{2008MNRAS.388.1335W} and \(b_{\rm EMU}(z) \equiv 2.32\) from \citet{2025PASA...42...62T}.
\end{itemize}
Whilst \citet{2025PASA...42...62T} and \citet{Saraf:EMUxDES} also use \(n_\mathrm{EMU}(z)\) models inferred from the Tiered Radio Extragalactic Continuum Simulation \citep[T-RECS;][]{2019MNRAS.482....2B}, we only consider SKADS as our baseline as the routine employed to generate \(w(\hat{\vec n})\) is intrinsically based on SKADS.

All theoretical predictions for the harmonic-space power spectrum are obtained using the `Core Cosmology Library' \citep[\texttt{CCL},\footnote{\url{https://ccl.readthedocs.io/en/latest/.}}][]{Chisari_2018}, assuming a fiducial \(\Lambda\)CDM cosmological model with parameters from \cite{2020A&A...641A...6P} and when computing the power spectrum, we modelled the nonlinear corrections with 
the \texttt{halofit} model \citep{SMith2003}.


\section{Power spectrum estimation}\label{sec:methodology}\label{sec:Clestimation}
We measure the harmonic-space power spectrum from the data using the pseudo-\(C_\ell\) formalism as implemented in \texttt{NaMaster}.\footnote{\url{https://github.com/LSSTDESC/NaMaster}.} We briefly describe the estimator below \citep[for a detailed presentation, see][]{2019MNRAS.484.4127A,2025JCAP...01..028W}. For the analysis of correlations between two different galaxy over-density maps, as is considered here, we deal with two scalar fields (labelled by \( A\) and \( B\)) defined on a masked sky. Their observed power  spectrum, \(\tilde{C}_\ell^{AB}\), is a biased estimator of the underlying power spectrum \(C_\ell^{AB}\), viz.\
\begin{equation}\label{eq:cls_true}
    \left\langle \tilde{C}_\ell^{AB} \right\rangle = \sum_{\ell'} M_{\ell \ell'}^{AB}\,C_{\ell'}^{AB} + N_\ell^{AB}\;,
\end{equation}
where \(N_\ell^{AB}\) is the noise bias (later discussed in \cref{ssec:shot_noise}) and \(M^{AB}_{\ell\ell'}\) is the coupling matrix. The latter depends only on the features of the mask(s).

For this, we can build an unbiased estimator \(\hat C_\ell^{AB}\) of the underlying power spectrum as 
\begin{equation}
    \hat{C}_\ell^{AB} = \sum_{\ell'} \left[\left(M_{\ell \ell'}^{AB}\right)^{-1}\right]_{\ell \ell'} \left( \tilde{C}_{\ell'}^{AB} - N_{\ell'}^{AB} \right)\;. \label{eq:clestimator}
\end{equation}
To further reduce the effect of mode-coupling, we compute the harmonic-space power spectrum in \(N_{\rm bp}=8\) band-powers, linearly spaced in \(\ell\) with \(\Delta\ell=100\) starting from \(\ell_{\min}=2\). This should be enough to mitigate effect of small covered area \citep{POLARBEAR_2019}. 
The choice of minimum angular scale included in the analysis follows the preliminary official forecasts by \citet{Blanchard-EP7}, who identified in \(\ell=750\) the most conservative scenario to avoid  nonlinearities. Given our band-power set-up, for us this corresponds to \(\ell_{\max}=801\). In the following, we shall also show how the results are stable against our choice of \(N_{\rm bp}\).

To estimate the data covariance, we use the analytical derivation of \cite{Garcia-Garcia_2019} and its implementation in {\tt NaMaster}. In particular, we assume that the two galaxy overdensity fields are random Gaussian variables and we correct for the mode coupling induced by the presence of the mask using the narrow kernel approximation. Given the small patch in the sky covered by the two surveys, we verified with simulations that this approximation is effective and sufficient to account for the limited survey area (\(f_{\rm sky}\approx0.5\text{\textperthousand}\)), thus regularising the otherwise singular, non-invertible coupling matrix (see \cref{apdx:cov}).

\subsection{Shot noise}\label{ssec:shot_noise}
When measuring the auto-spectrum, the mode-coupled noise bias for the auto-correlation is \(\smash{N_\ell^{AA}}\). Source counting is a Poisson process \citep[e.g.][]{Tessore:2025auu}, thus, \(\smash{N_\ell^{AA}}\) is usually related to the Poisson shot noise, defined as \(N_\ell^{AA} = \Omega_p/\bar N_w\),
where the mask is averaged over the sky, \(\Omega_p\) is the solid angle of each \texttt{HEALPix} pixel \(p\). 

\Euclid galaxies are expected to form a Poissonian sample. Whilst the \Euclid galaxy distribution is inherently clustered, its shot noise term is expected to follow the relation above due to the high source density and uncorrelated detection of individual objects. On the other hand, for high-resolution and high-sensitivity radio surveys, this Poissonian assumption may break down due to overdispersion from multi-component sources.\footnote{Here, `multi-component sources' refers specifically to the extended radio morphology (e.g., core, jets, and lobes) of a single physical galaxy, which can be spatially resolved and catalogued as multiple distinct sources. This differs from optical multi-components, which typically refer to internal star-forming clumps within a much smaller physical scale, usually identified as a single source. Overdispersion occurs when these widely separated radio components are counted individually, artificially boosting the measured angular power spectrum at small scales.} This non-Poissonianity stems from a spurious small-scale clustering signal caused by the proximity of multiple components belonging to a single physical source. In such cases, the negative binomial distribution \citep{johnson2005} often provides a more accurate representation of the counts and their associated noise \citep{Pashapour:countsincellstats}. We, therefore, examine whether the EMU data also follow a Poissonian behaviour.

A simple diagnostic for Poissonianity is the clustering parameter \(n_\mathrm{c}\) \citep{1980lssu.book.....P}, defined as the ratio of the variance to the mean of the weighted number counts in pixels \(p\) over the observed footprint, viz.\
\begin{equation}
    n_\mathrm{c} = \frac{\left\langle\left[N_w(\hat{\vec n}) - \bar N_w\right]^2\right\rangle}{\bar N_w}\;.
\end{equation}
For a truly Poissonian sample, where the mean and variance are equal, we expect \(n_\mathrm{c} = 1\). A value significantly higher than unity implies super-Poissonian behaviour. In our case, we find the \selavy\ island sample to be almost perfectly Poissonian, with \(n_\mathrm{c} = 1.06\) whilst for the \pybdsf\ source catalogue, we measure a slightly higher value of \(n_\mathrm{c} = 1.28\). We do not fully understand why the \pybdsf\ catalogue gives a larger \(n_\mathrm{c}\) value. One reason could be that the \pybdsf\ catalogue extends into areas masked out in the \selavy\ catalogue (cf. \cref{fig:footprints}). When restricting the \pybdsf\ map to the \selavy\ footprint, we measure \(n_\mathrm{c} = 1.20\). Although this is still lower than any values reported for LoTSS samples by \citet{Pashapour:countsincellstats}, it indicates a mild deviation from Poissonianity for \pybdsf. For the purposes of this analysis, we approximate both the \selavy\ and \pybdsf\ catalogues with Poissonian shot noise, acknowledging that we will revisit this assumption in future work (see also \cref{sec:cic}).

Given that \Euclid and EMU galaxy distributions are approximately Poissonian, we can model the selection of EMU galaxies (in the absence of multi-component sources) from the \Euclid\ parent sample as a thinning process. Thinning a Poisson process results in another Poisson process, whose shot noise contribution can be estimated as \citep[see also][]{2016MNRAS.463.3674H,2024OJAp....7E..15U}
\begin{equation}
    N_\ell^{AB} = \frac{\bar n_{A\wedge B}}{\bar n_A\,\bar n_B}\;,
\end{equation}
where \(\bar n_{A\wedge B}\) represents the angular number density of objects in common between the two catalogues. It is easy to see that the expression above resorts to the familiar \(\smash{N_\ell^{AB}=\delta_{AB}^{\rm(K)}/\bar n_A}\) for disjoint galaxy populations, with \(\smash{\delta^{\rm(K)}}\) the Kronecker-delta symbol.

In our case, since \Euclid has a significantly higher source density, we introduce the parameter \(f_\wedge\coloneqq\bar n_{\mathrm{EMU\wedge Euc}}/\bar n_\mathrm{EMU}\), which indicates the ratio of EMU sources that have a \Euclid counterpart. Considering that \(f_\wedge \in [0,1]\), we can establish an upper limit on the Poissonian cross-noise, namely
\begin{equation}
    \frac{f_\wedge}{\bar n_\mathrm{Euc}} \leq \frac1{\bar n_\mathrm{Euc}} \approx \begin{cases}
        8.0 \times 10^{-9}\,\mathrm{sr}^{-1}& (\texttt{pgal0.90}) \\
        4.4 \times 10^{-9}\,\mathrm{sr}^{-1}& (\texttt{pstar0.05})
    \end{cases}\;.
\end{equation}
Using simple positional cross-matching presented in \cref{apdx:xmatch}, we estimate \(f_\wedge \lesssim 0.3\), which remains to be confirmed by other EMU projects undertaking more sophisticated cross-matching analyses. In any case, we shall demonstrate that the cross-noise value is at least an order of magnitude lower than the expected cross-correlation signal within our fiducial multipole range, making it subdominant.

\section{Results}
\label{sec:results}
We start by presenting the measurement of the EMU-\Euclid cross-spectrum, moving then to assessing the statistical significance of the detection. We shall then put bounds on the cross-shot noise. Finally, we show how this combined data set can be used to learn about the redshift distribution of EMU sources.

\subsection{Measurement of the EMU-\Euclid cross-spectrum}
\label{ssec:results.measured_cls}
\begin{figure*}
    \centering
    \includegraphics[width = 0.9\linewidth]{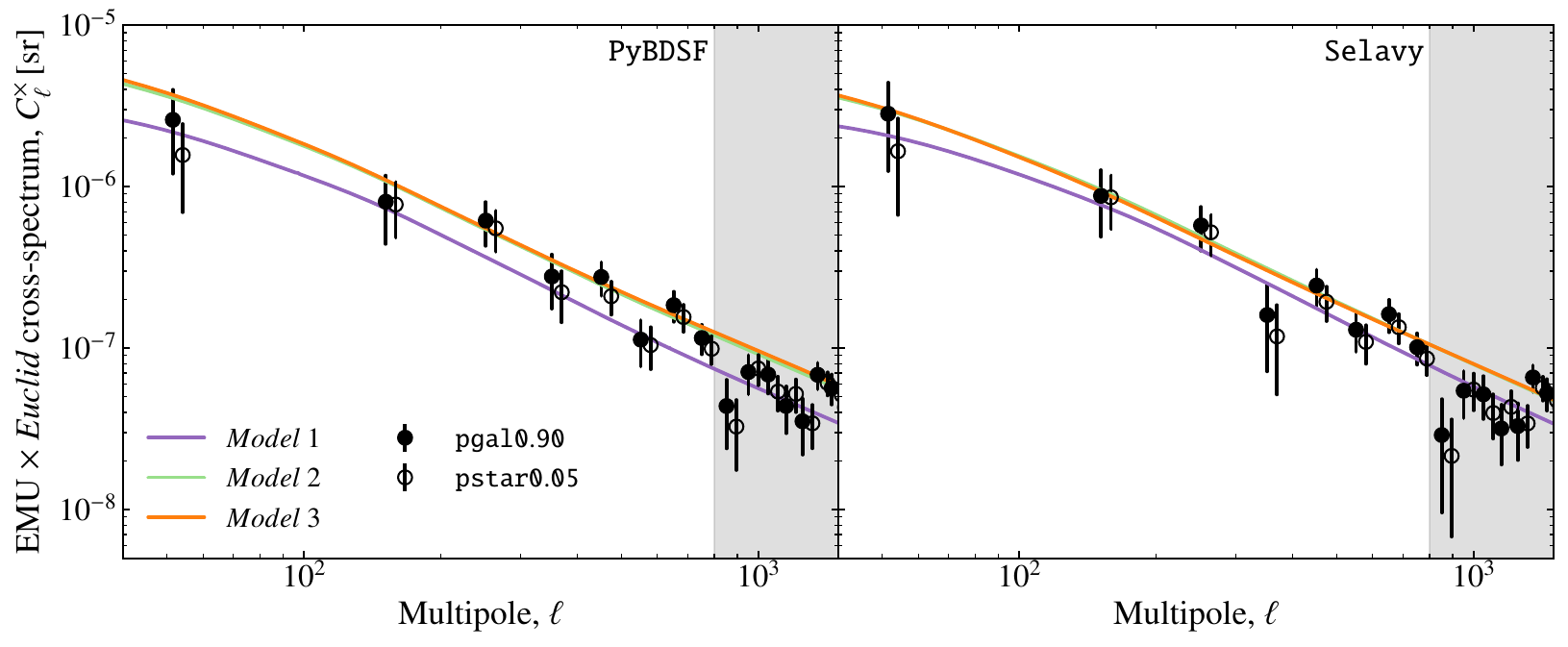}
    \caption{
    Measured harmonic-space cross-spectra between \Euclid and EMU. The black markers represent the measured cross-spectra for the two EMU baseline maps (\textit{left panel} \pybdsf~ and \textit{right panel} \selavy) combined with the two \Euclid maps. Coloured lines show the theory predictions for the cross spectrum using the fiducial redshift distributions and bias models discussed in \cref{ssec:theory_met.theory}. Note that the grey shadowed areas represent the multipoles excluded from our analysis. In each panel, full and empty markers are for \pgal and \pstar~ respectively.}
    \label{fig:cls_fidu}
\end{figure*}

Using \cref{eq:cls_true}, we estimate the harmonic-space cross-correlation pseudo-\(C_\ell\) band-powers between EMU and \Euclid maps (for simplicity, \(C^\times_\ell\) hereafter). The results, for the various combinations of EMU maps (\selavy\ and \pybdsf) and \Euclid maps (\pgal\ and \pstar), are shown in \cref{fig:cls_fidu}, with different markers and error bars. In the same figure, we over-plot the theoretical cross-spectra, computed with \cref{eq:Cl_Limber}, in the fiducial \Planck\ cosmology (cf. \cref{ssec:theory_met.theory}) and with the various choices of redshift distributions and bias also described in \cref{ssec:theory_met.theory} (coloured curves).

Notably, the cross-spectra are in agreement within \(1\,\sigma\) error bars across the entire range of multipoles. In the context of EMU, this outcome is in line with the results of \citet{2025PASA...42...62T}, in which the cross-correlation was found to be significantly more robust against spurious signals that may be related to the systematic effects in observations or, in our case, to uncertainties in the source-finding algorithm.
Motivated by the remarkable qualitative agreement between the model and the data,
we choose to parameterise our model by a single amplitude parameter \(A_b\), which will allow for a rigid rescaling of the curves so that they can best fit the data points.

To proceed, we construct a chi-squared statistic for our data. For a general parameter set \(\vec\vartheta\), we write
\begin{equation}
    \chi^2(\vec\vartheta)=\left[\vec d-\vec m(\vec\vartheta)\right]^\dagger\,\tens\Sigma^{-1}\,\left[\vec d-\vec m(\vec\vartheta)\right]\;,\label{eq:X2}
\end{equation}
where \(\vec d\) is the data vector, i.e., the mask-deconvolved pseudo-\(C_\ell\) band-powers, \(\vec m\) is the theoretical model, \(\tens\Sigma\) is the covariance matrix, and we denote matrix inversion and transposition respectively by \(-1\) and \(\dagger\). For our single, one-parameter amplitude model, \(\vec\vartheta=\{A_b\}\) and \(\vec m=A_b\,\vec t\), with \(\vec t\) the fiducial theory vector from \cref{eq:Cl_Limber}, averaged over the same band-powers as the data. The minimisation of such a chi-squared has a known analytical solution, and the best-fit value (marked by an asterisk) and standard deviation on the parameter \(A_b\) read
\begin{align}
    A_b^\ast &= \frac{\vec t^\dagger\,\tens\Sigma^{-1}\,\vec d}{\vec t^\dagger\,\tens\Sigma^{-1}\,\vec t}\;,\label{eq:Ab}\\
    \sigma_{A_b^\ast} &= \frac{1}{\sqrt{\vec t^\dagger\,\tens\Sigma^{-1}\,\vec t}}\;.\label{eq:sigma_Ab}
\end{align}

\begin{figure}
    \centering
    \includegraphics[width=\linewidth]{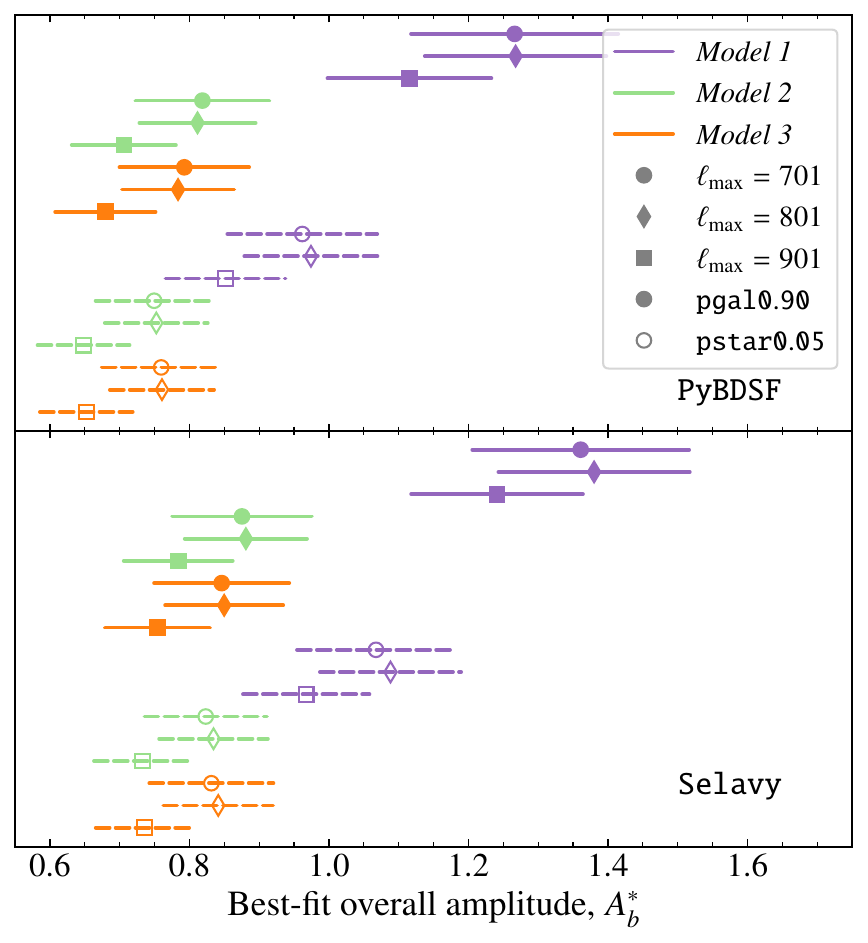}
    \caption{Best-fit overall amplitude, \(A^{\ast}_b\), for the one-parameter fit, as in \cref{eq:Ab,eq:sigma_Ab}. The figure demonstrates the high degree of internal consistency among all modelling choices (colour and line-style code as in \cref{fig:cls_fidu}) and data subsets (upper/lower panel for \pybdsf/\selavy; filled/empty markers for \pgal/\pstar). The main conclusion is that the best-fit \(A^{\ast}_b\) is highly robust against the choice of EMU catalogue and \(\ell_{\max}\) scale cut (marker shapes), allowing us to interpret \(A^{\ast}_b\) as a measure of the overall rescaling required for the FS2 simulation's \Euclid bias prescription.}
    \label{fig:Ab}
\end{figure}

The best-fit values of \(A_b\) and corresponding \(1\,\sigma\) error bars for all the combinations of data sets and modelling choices are shown in \cref{fig:Ab}. Following the same colour-code as \cref{fig:cls_fidu}, different colours mark the various modelling of the EMU signal, with violet for the redshift distribution and bias from \citet{Saraf:EMUxDES}, green and orange for SKADS from \citet{2008MNRAS.388.1335W} combined with redshift-dependent or constant bias, as in \citet{2025PASA...42...62T}, respectively. Regarding the data, the two panels distinguish between EMU maps constructed using either the \selavy\ or the \pybdsf\ source finders, whereas filled and empty markers differentiate the \pgal\ from \pstar\ \Euclid catalogues. Finally, to validate the robustness of our results against our choice of scale cuts, we show with different markers the results of the fit when shrinking/enlarging the multipole range. 
In particular, our baseline choice of \(8\) band-powers in the multipole interval \(\ell\in[2,801]\) is represented by diamond markers, whereas the result of removing or adding one additional band-power with the same step \(\Delta\ell=100\) is shown with circle or square markers, respectively. 

It is immediately apparent that the fits show an excellent degree of internal consistency. On the EMU side, our analysis is in full agreement with \citet{2025PASA...42...62T}, who found no definite preference between a constant bias \(2.32\) and a constant-amplitude bias \(1.81/D(z)\) when using the \(n(z)\) from the SKADS simulations \citep[][see light-blue curves in \cref{fig:dndz_bias}]{2008MNRAS.388.1335W}, regardless of the source finder employed for EMU galaxies.\footnote{This is also in line with \citet{2024A&A...681A.105N}'s LoTSS analysis who could only disfavour a constant bias once auto- and cross-spectra are combined.} However, the more recent analysis by \citet{Saraf:EMUxDES}, who cross-correlated the EMU Pilot Survey 1 with optical galaxies from DES, reported a statistically significant preference for a radio-continuum source distribution peaked at a higher redshift than what is predicted by SKADS (see light- vs dark-blue curves in the upper panel of \cref{fig:dndz_bias}). Adopting this source distribution (\textit{Model 1}) predicts a lower cross-spectrum, requiring a significantly larger value for \(A_b\) when cross-correlated with \pgal.

In light of this, 
it is worth asking ourselves what the meaning of \(A_b\) is. The two adopted \(n_\mathrm{Euc}(z)\) are obtained from the \Euclid catalogues. We have major remaining modelling uncertainties such as the \(n_\mathrm{EMU}(z)\) and the different \(b_\mathrm{EMU}(z)\), which are fits to the data but with different assumptions, as well as the bias of \Euclid galaxies, which we have modelled using a polynomial fit to the FS2 simulation. Given that the shape of the theoretical cross-spectra is consistent with the data points (see \cref{fig:cls_fidu}), it is hence natural to interpret \(A_b\) as an overall rescaling of the FS2 bias prescription. It should not come as a surprise that \pgal\ and \pstar\ best-fits for \(A_b\) follow two separate trends, for as long as they agree within standard deviations. Indeed, the two choices of flag cuts in the \Euclid catalogue result in different source redshift distributions (light- and dark-red curves in \cref{fig:dndz_bias}, upper panel). Likely, they represent two subsamples of the galaxy population observable by \Euclid, and it is natural to expect that they have different biases.


We quantify the significance of the detection of the cross-correlation signal as
\begin{equation}
\label{eq:significance}
    \sigma\coloneqq A_b^{\ast}/\sigma_{A_b^{\ast}} .
\end{equation}
The results are listed in \cref{tab:phz_chi2} when considering different values of \(\ell_{\rm max}\) used in our analysis.
From \cref{tab:phz_chi2}, we quote a high detection significance for all the different combinations of \Euclid and EMU maps, going from a minimum of about \(8.7\,\sigma\) for \(\ell_{\rm max} = 701\) and {\it Model 1} to a maximum of about \(10.8\,\sigma\) for \(\ell_{\rm max} = 801\).

\begin{table}
    \centering
    \caption{\textit{First column:} Maximum multipole included in the analysis. \textit{Second and third columns: } \Euclid and EMU data sets employed, respectively. The theoretical models described in \cref{ssec:theory_met.theory}. \textit{Fourth to last columns:} Detection significance \(\sigma\) (Eq.\ \ref{eq:significance}) of the cross-spectrum for the three models adopted for the radio-continuum sources (see \cref{ssec:theory_met.theory}).}
    \begin{tabularx}{\columnwidth}{X cc ccc}
    \toprule
     & & & \multicolumn{3}{c}{Detection significance, \(\sigma\)} \\
     \cline{4-6} \\ [-2ex]
    \(\ell_{\rm max}\) & \Euclid & EMU & {\it Model 1} & {\it Model 2} & {\it Model 3} \\
    \midrule\midrule
     \multirow{4}{*}{701} & \multirow{2}{*}{\pgal} & \pybdsf & 8.76 & 8.74 & 8.74 \\
      & & \selavy & 8.53 & 8.53 & 8.53 \\
      & \multirow{2}{*}{\pstar} & \pybdsf & 9.43& 9.37 & 9.35 \\
      & & \selavy & 8.95 & 8.94 & 8.93 \\
      \midrule
     \multirow{4}{*}{801} & \multirow{2}{*}{\pgal} & \pybdsf & 10.06 & 10.05 & 10.05 \\
      & & \selavy & 9.74 & 9.74 & 9.73 \\
      & \multirow{2}{*}{\pstar} & \pybdsf & 10.75 & 10.71 & 10.68 \\
      & & \selavy & 10.20 & 10.19 & 10.18 \\
      \midrule
     \multirow{4}{*}{901} & \multirow{2}{*}{\pgal} & \pybdsf & 10.08& 10.04& 10.02 \\
      & & \selavy &9.49 &9.46 & 9.44\\
      & \multirow{2}{*}{\pstar} & \pybdsf &10.67 &10.58 & 10.55\\
      & & \selavy & 9.90& 9.83& 9.81 \\
    \bottomrule
    \end{tabularx}
\label{tab:phz_chi2}
\end{table}

\subsection{Including cross-shot noise}\label{sec:Xnoiseest}
\begin{table*}
\centering
\caption{Best-fit values for the overall amplitude \(A^\ast_b\) and \(C_{\rm shot}^{\times\ast}\) from the analytical \(\chi^2\) minimisation from \cref{eq:theta_2D}. {\it First and second columns: } \Euclid and EMU data sets employed, respectively. {\it Third column}: value of \(\ell_{\rm max}\). {\it Fourth and fifth columns}: values of \(A_{b}^{\ast}\) with the \(1\,\sigma\) error. {\it Sixth and seventh columns}: values of cross-shot noise with the \(1\,\sigma\) error. {\it Eighth and ninth columns}: value of the reduced \(\chi^2\) where the number of degrees of freedom is the \({\rm d.o.f.} = 8\,\text{(number of band powers)}-2\,\text{(number of free parameters)}\).}
\begin{tabular}{llc ccc ccc cc}
\toprule
\Euclid & EMU & \(\ell_{\rm max}\) & \multicolumn{2}{c}{\(A_b^\ast\pm\sigma_{A_b^\ast}\)} && \multicolumn{2}{c}{\(C^{\times\ast}_{\rm shot}\pm \sigma_{C^{\times\ast}_{\rm shot}}\,[10^{-8}\,\mathrm{sr}]\)} && \multicolumn{2}{c}{\(\chi^2/{\rm d.o.f.}\)} \\ [0.75ex]
\cline{4-5}\cline{7-8}\cline{10-11}
&&& \pybdsf & \selavy && \pybdsf & \selavy && \pybdsf & \selavy\\
\midrule\midrule
\multirow{3}{*}{\pgal} &  {\it Model 1} & 801 & 1.26\(\pm\)0.30 & 1.18\(\pm\)0.29 && 1.65\(\pm\)3.56 & 1.12\(\pm\)3.45 && 0.84 & 0.68 \\
 & {\it Model 2} & 801 & 0.82\(\pm\)0.20 & 0.79\(\pm\)0.19 && 1.23\(\pm\)3.67 & 0.48\(\pm\)3.59 && 0.89 & 0.68 \\
 & {\it Model 3} & 801 & 0.80\(\pm\)0.19 & 0.77\(\pm\)0.19 && 1.15\(\pm\)3.70 & 0.32\(\pm\)3.63 && 0.91 & 0.69 \\
 
\midrule
\multirow{3}{*}{\pstar} & {\it Model 1}  & 801 & 0.94\(\pm\)0.21 & 0.86\(\pm\)0.21 && 2.16\(\pm\)2.79 & 1.61\(\pm\)2.69 && 0.79 & 1.01 \\
 & {\it Model 2} & 801 & 0.72\(\pm\)0.17 & 0.68\(\pm\)0.16 && 2.29\(\pm\)2.81 & 1.39\(\pm\)2.76 && 0.93 & 1.06 \\
 & {\it Model 3} & 801 & 0.72\(\pm\)0.17 & 0.69\(\pm\)0.17 && 2.40\(\pm\)2.82 & 1.37\(\pm\)2.78 && 1.00 & 1.10 \\
\bottomrule
\end{tabular}
\label{tab:ab_crossSn}
\end{table*}

Given the relatively high value of \snr\ listed in \cref{tab:phz_chi2} (up to \(\snr=10.75\) for \pybdsf\ at \(\ell_{\rm max} = 801\)), we can investigate the joint constraints on both the overall amplitude \(A_b\) and the cross-shot noise, \(C^\times_{\rm shot}\). In particular, if we add a constant shot-noise parameter to our model, i.e.\
\begin{equation}
    \vec m=A_b\,\vec t+C^\times_{\rm shot}\;,
\end{equation}
we can still minimise analytically the chi-squared of \cref{eq:X2}. If the parameter set is \(\vec\vartheta=\{A_b,C^\times_{\rm shot}\}\), the best-fit estimates for the parameters are given by
\begin{equation}
\label{eq:theta_2D}
    \vec\vartheta^\ast=(\tens T^\dagger\,\tens\Sigma^{-1}\,\tens T)^{-1}\,\tens T^\dagger\,\tens\Sigma^{-1}\,\vec d\;,
\end{equation}
with \(\tens T=(\vec t\quad\vec1)^\dagger\).
The uncertainties on the two parameters can then be obtained from the parameter covariance matrix, \(\tens T^\dagger\,\tens\Sigma^{-1}\,\tens T\).
We note that this reduces to \cref{eq:Ab,eq:sigma_Ab} when minimising for one parameter alone.
The resulting values of \(A^{\ast}_b\) and a \(1\,\sigma\) upper bound on \(C_{\rm shot}^{\times\ast}\) are listed in \cref{tab:ab_crossSn} for the benchmark value of \(\ell_{\rm max} = 801\). Results on the overall amplitude are in agreement, within \(1 \,\sigma\) error, with the estimates obtained by minimising the 1D \(\chi^2\) of Eq.\ (\ref{eq:X2}, see \cref{fig:Ab}). In this case, considering a larger parameter space given by the presence of two free parameters of the model, the error bars on \(A^{\ast}_{b}\) are larger with respect to the results of \cref{fig:Ab} up to a factor of about \(2\). In any case, we obtain a cross-shot noise \(C^\times_\mathrm{shot}\) consistent with zero, with \(1\,\sigma\) bounds of an order of magnitude higher than the one computed analytically in \cref{ssec:shot_noise}. Furthermore, we do not see any improvement in terms of \(\chi^2/{\rm d.o.f.}\), meaning that we do not have to worry further about the shot-noise term. In most cases, \(\chi^2/{\rm d.o.f.}\) actually becomes slightly worse in the joint minimisation. For instance, using \pgal\ and \selavy\ assuming {\it Model 1} in the single parameter $A_b$ fit provides \(\chi^2/{\rm d.o.f.} = 0.60\), whereas \pstar\ and \selavy\ assuming {\it Model 3} results in \(\chi^2/{\rm d.o.f.} = 0.97\), to name the most extreme cases of \autoref{tab:ab_crossSn} for comparison.  

\begin{figure}[h]
    \centering
    \includegraphics[width=\linewidth]{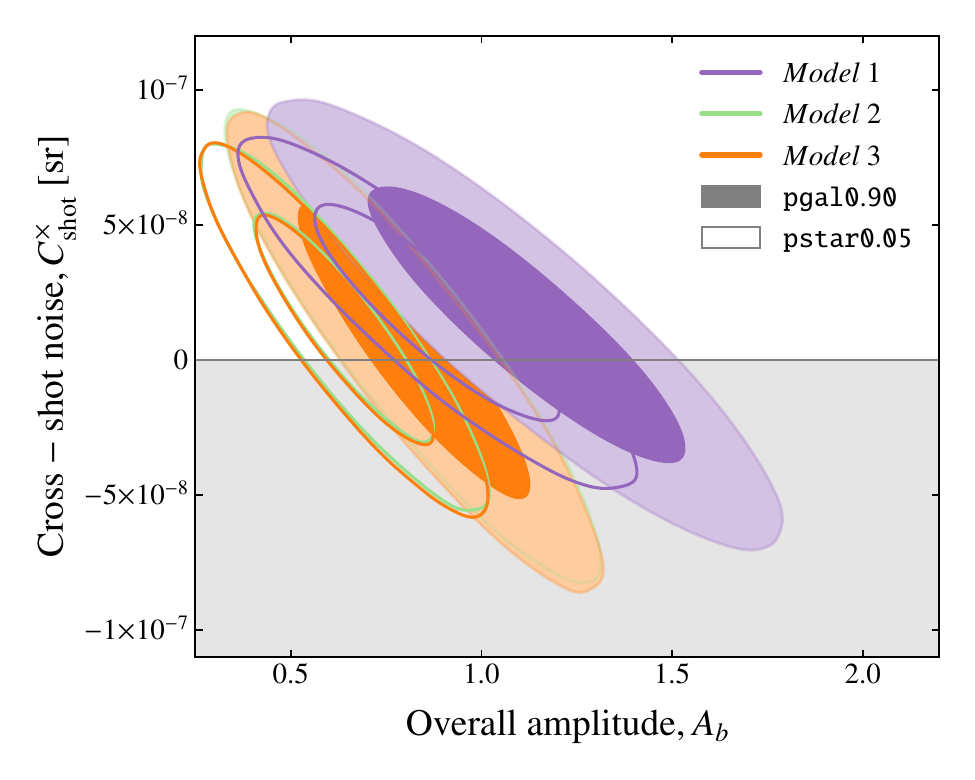}
    \caption{\(A_b\)-\(C^\times_\mathrm{shot}\) posterior contours corresponding to the results in 
    \cref{tab:ab_crossSn}, i.e., before applying a prior on \(C^\times_\mathrm{shot}\). We only present results for \pybdsf\ here for clarity. The parameter correlations obtained with \selavy\ are consistent with the contours presented here. The grey shadowed area represents the region of parameter space excluded by the prior on \(C_{\rm shot}^{    \times}\) that is, for unphysical negative values of the cross-shot noise.}
    \label{fig:AbCshot_2dcontours}
\end{figure}

The \(A_b\)-\(C^\times_\mathrm{shot}\) contours in \cref{fig:AbCshot_2dcontours} reveal a strong degeneracy between the bias amplitude, \(A_b\), and the cross-shot noise, \(C^\times_\mathrm{shot}\). Consistent across both the \(\pybdsf\) and \(\selavy\) results, the \(C^\times_\mathrm{shot}\) posterior region extends significantly into negative values. This extension confirms that the data are compatible with zero shot noise.

Whilst the maximum a posteriori (MAP) value for \(C^\times_\mathrm{shot}\) is technically positive, this is irrelevant given the large error bars. Imposing the physical prior \(C^\times_\mathrm{shot} \geq 0\) shifts the \(A_b\) posterior means toward lower values; this is a pure projection effect due to truncating the degenerate parameter space. Critically, the upper bounds on \(C^\times_\mathrm{shot}\) remain robust, confirming that the data do not favour a large positive shot noise.

\subsection{Reconstruction of the EMU source redshift distribution}\label{sec:Nzfit}

Whilst fitting for an overall amplitude \(A_b\) provides a robust measure of the signal significance and consistency, a primary goal of cross-correlation analyses is to leverage the tomographic information from the well-characterised survey (\Euclid) to constrain the unknown redshift distribution of the other (EMU). Having established a high-significance detection, we now move beyond the fixed \(n_{\rm EMU}(z)\) models used in the previous sections, and instead allow the shape of the EMU redshift distribution to be a free function, constrained directly by our measured cross-spectrum.

To achieve this, we remodel the EMU redshift distribution using three distinct parameterisations, which allows us to test the robustness of our results against different assumptions about the underlying functional form. In each case, we perform a fit to the measured cross-spectrum by minimising the \(\chi^2\) statistic as defined in \cref{eq:X2}, now as a function of the parameters describing the EMU source redshift distribution, instead of \(A_b\). 

To better constrain the redshift evolution of \(n_{\rm EMU}(z)\), we utilise the \(\Euclid\) \texttt{pstar0.05} sample, subdividing it into three tomographic redshift bins (with distributions shown in \cref{fig:tomography}). We then measure the cross-spectra between these three \(\Euclid\) tomographic bins and the \(\text{EMU}\) \texttt{pybdsf} sample.

We choose to fix the \(\text{EMU}\) galaxy bias to the functional form \(b_{\rm EMU}(z) = 1.81/D(z)\) from the baseline analysis by \citet{Saraf:EMUxDES}. This model choice is strongly supported by the data combination used: it provides the best agreement with the fiducial bias assumption, yielding \(A_b = 0.94 \pm 0.21\) (see \cref{tab:ab_crossSn}), particularly for \textit{Model 1}. Fixing the bias is necessary because we found that joint fitting for both \(n_{\rm EMU}(z)\) and \(b_{\rm EMU}(z)\) simultaneously leads to unstable parameter constraints and poorly defined confidence regions.

We acknowledge a crucial caveat: the fixed \(b_{\rm EMU}(z)\) used here was conditioned on the specific \(n_{\rm EMU}(z)\) model used in the original \citet{Saraf:EMUxDES} work. Consequently, varying the \(n_{\rm EMU}(z)\) functional form in our current fits whilst keeping the bias fixed may introduce a model-dependent shift in the inferred peak or shape of the final \(n_{\rm EMU}(z)\). A proper, self-consistent joint determination of \(n_{\rm EMU}(z)\) and \(b_{\rm EMU}(z)\) requires information from the \(\text{EMU}\) auto-spectrum. To reliably use the auto-spectrum, we must first conduct more thorough tests on source-finding uncertainty, scale cuts, and other systematic factors, which are planned for future analyses and are beyond the scope of this publication.

Moreover, we note that the inclusion of the \Euclid auto-spectrum also plays a crucial role in lifting degeneracies with the galaxies bias therefore providing more stringent constraints.
The aim of this pathfinder analysis is, however, to showcase how EMU-\Euclid\ cross-correlations will be able to constrain the redshift distribution of radio sources with high fidelity. 

We consider the three parameterisations described below:

\begin{itemize}

\item{\bf Baseline parameterisation.} We adopt the functional form used for LoTSS DR1 \citep[][and also by \citealt{Saraf:EMUxDES}]{2021MNRAS.502..876A}, which reads 
    \begin{equation}
    n_{\rm EMU}(z) \propto \frac{(z/z_{0})^{2}}{1+(z/z_{0})^{2}}\,\frac{1}{1+(z/z_{\text{tail}})^{\gamma}}\;.
    \label{eq:model_nz_lotssdr1}
\end{equation}

\item{\bf Physically motivated model.} As a second approach, we employ the more complex physically motivated parameterisation developed for LoTSS DR2 \citep{2024A&A...681A.105N,2024MNRAS.527.6540H}. This model comes from the fact that the radio source population is a mixture of SFG and AGN, with a certain ratio \(q\). 
The functional form combines two main terms: an exponential cut-off and a power law. The exponential term, which dominates the SFG population at low redshift, is motivated by the Schechter function typically used to model the radio luminosity functions of SFGs. The power law, characterising the AGN population that extends to higher redshifts, arises from the typical double power-law approximation for AGN radio luminosity functions. The functional form is given by
\begin{equation}
    n_{\rm EMU}(z) \propto \frac{z^2}{1 + z}\,\left[\exp\left(-\frac{z}{z_0}\right) + \frac{q^2}{\left(1 + z\right)^a}\right]\;.
\end{equation}
The overall \(z^2/(1 + z)\) pre-factor accounts for the expected growth in source numbers, proportional to the volume, at low redshifts in a de Sitter cosmology.\footnote{A spatially flat universe dominated by the cosmological constant without ordinary matter as a zeroth-order approximation to the low-\(z\) \(\Lambda\)CDM Universe.} At higher redshifts, the sample's flux density limitation becomes the dominant factor, making the form of each population's luminosity function critical.

\item{\bf Cubic-spline interpolation.} Finally, to make minimal assumptions about the shape of the EMU source redshift distribution, we use a non-parametric approach. We model \(n_{\rm EMU}(z)\) as a cubic spline curve defined by a series of nodes at fixed redshift intervals \(\Delta z = 0.75\). We chose this interval to cover the \Euclid redshift range with three nodes maintaining the number of free parameters equal to those in the previous cases. The amplitudes of the \(n_{\rm EMU}(z)\) at these nodes are left as free parameters in the fit. This method is highly flexible, allowing the data to determine the shape of the distribution, providing a powerful check for any features not captured by the fixed functional forms.
\end{itemize}
\begin{table}[h]
\centering
\caption{Prior distributions for different parameterisation of the EMU source redshift distribution model, with \(\mathcal{U}(a, b)\) representing the uniform distribution in the interval \([a,b]\).}
\begin{tabular}{ l  c }
\toprule
\multicolumn{2}{c }{\textbf{{Baseline parameterisation}}}\\
 \midrule
\(z_0\) & \(\mathcal{U}(0, 15)\) \\ 
\(z_{\rm tail}\) & \(\mathcal{U}(0, 15)\) \\ 
\(\gamma\) & \(\mathcal{U}(0, 20)\) \\ 
\midrule
 \midrule
\multicolumn{2}{ c }{\textbf{{\small Physically motivated model}}}\\
 \midrule
 \(z_0\) & \(\mathcal{U}(0, 6)\) \\ 
 \(q\) & \(\mathcal{U}(0, \infty)\) \\
  \(a\) & \(\mathcal{U}(0, 20)\) \\
 \midrule
 \midrule
   \multicolumn{2}{ c }{\bf Cubic-spline interpolation}\\
 \midrule
   \(p_\mathrm{EMU}(z = 0.75)\) & \(\mathcal{U}(0, 4)\) \\
   \(p_\mathrm{EMU}(z = 1.5)\) & \(\mathcal{U}(0, 2)\) \\
   \(p_\mathrm{EMU}(z = 2.25)\) & \(\mathcal{U}(0, 1)\) \\
    \bottomrule
\end{tabular}

\label{tab:priors}    
\end{table}
\begin{table*}
\centering
\caption{Posterior mean values of the \(p_\mathrm{EMU}(z)\) parameterisations considered in \cref{sec:Nzfit} with their corresponding negative double log-posterior \(\smash{\chi_\mathrm{MAP}^2 + \xi_\mathrm{MAP}^2}\), negative double log-prior \(\smash{\xi_\mathrm{MAP}^2}\), negative double log-likelihood \(\smash{\chi_\mathrm{MAP}^2}\), and goodness-of-fit \(\smash{\chi_\mathrm{MAP}^2/\mathrm{d.o.f.}}\)
We report a lower bound on \(p_\mathrm{EMU}(z=2.25)\) as its upper bound is set by the prior.}
\begin{tabular}{l r @{\hspace{0.25em}} S[table-format=1.3, table-align-text-after=false] 
 @{\hspace{2.5em}} 
 r @{\hspace{0.25em}} S[table-format=1.2, table-align-text-after=false] 
 @{\hspace{2.5em}} 
 r @{\hspace{0.25em}} S[table-format=2.1, table-align-text-after=false] 
@{\hspace{2.5em}} 
 cc cc }
\toprule
Parameterisation & \multicolumn{6}{c}{Posterior mean} & \(\chi_\mathrm{MAP}^2 + \xi_\mathrm{MAP}^2\) & \(\xi_\mathrm{MAP} ^ 2\) & \(\chi_\mathrm{MAP}^2\) & \(\chi_\mathrm{MAP}^2/\mathrm{d.o.f}\) \\
\midrule
Baseline & \(z_0=\) & 0.184\(^{+0.050}_{-0.031}\) & \(z_\mathrm{tail}=\) & 2.67\(^{+0.25}_{-0.23}\) & \(\gamma=\) & 9.7\(^{+7.9}_{-1.2}\) & \(27.0\) & \(5.98\) & \(21.0\) & \(1.17\) \\
Physically motivated & \(z_0=\) & 0.555 {\(\pm 0.025\)} & \(q=\) & 4.3 {\(\pm 1.7\)} & \(a=\) & 14.2 {\(\pm 4.1\)} & \(16.6\) & \(0\) & \(16.6\) & \(0.92\) \\
Cubic spline at \(z=\) & \(0.75:\) & 1.74\(^{+0.62}_{-0.51}\) & \(1.50:\) & 1.32\(^{+0.48}_{-0.26}\) & \(2.25\) & >0.667 & \(18.1\) & \(0\) & \(18.1\) & \(1.00\) \\
\bottomrule
\end{tabular}
\label{tab:Nz_maps}
\end{table*}

For each of these methodologies, we fit for the free parameters of the \(n_{\rm EMU}(z)\) whilst keeping the EMU bias model fixed. We normalise all distributions over the range \(z\in[0,6)\),
adding a penalty term for the first two parameterisations at the prior level, ensuring that \(p_{\rm EMU}(z = 6) < 10^{-5}\). This condition is naturally ensured in the cubic spline as we fix the nodes \(p_\mathrm{EMU}(z = 0) = p_\mathrm{EMU}(z = 6) = 0\) whilst intervals with negative spline realisations are interpreted as zero. 
The priors used in this analysis are summarised in \cref{tab:priors} for the different models. Note that for the baseline parametrisation, 
we furthermore apply an \(L^2\) penalty on the second derivative of \(n_\mathrm{EMU}(z)\) to avoid top-hat-like redshift distributions, with true top-hat distributions having infinite second derivatives. We do so in the form of a prior \(\mathcal{\Pi}(z_0, z_\mathrm{tail}, \gamma)\propto \exp(-\xi^2/2)\) with
\begin{equation}
    \xi^2 = 10\,\int_0^6 \de z\,\left|\frac{\de^2p_\mathrm{EMU}(z)}{\de z^2}\right|^2\;.
\end{equation}

\begin{figure}
\centering
\includegraphics[width = \columnwidth]{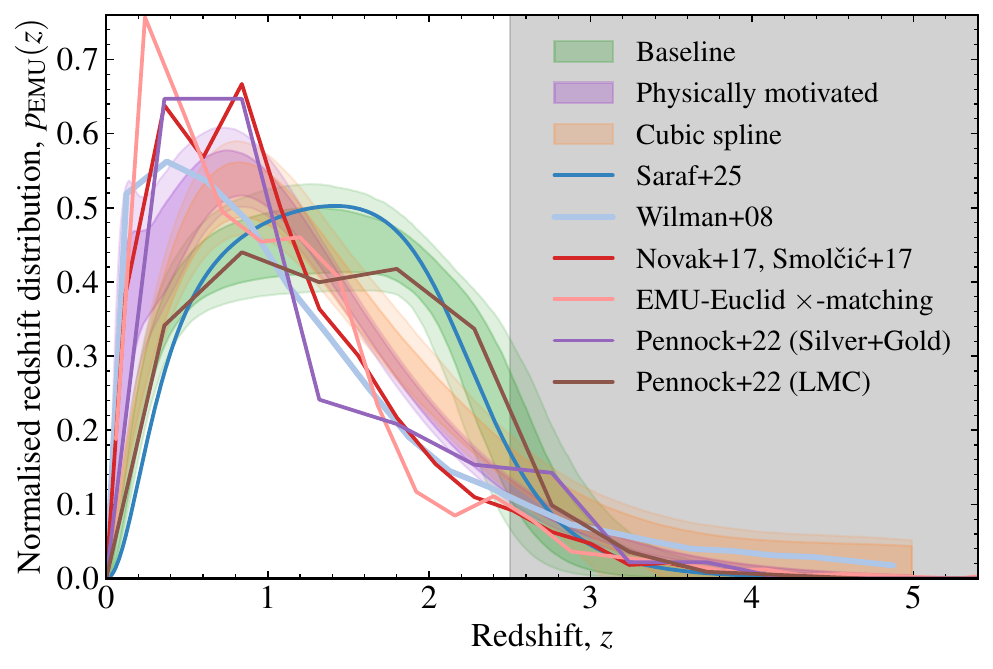}
\caption{Constraints on the EMU source redshift distribution, \(p_\mathrm{EMU}(z)\), from fitting the EMU-\Euclid cross-spectrum. The shaded bands show the \(1\) and \(2\,\sigma\) confidence intervals for the three different parameterisations: the fiducial function from \citet[][green]{2021MNRAS.502..876A,Saraf:EMUxDES}, the semi-analytic model (violet), and the non-parametric cubic spline fit (orange). For comparison, \cite{Saraf:EMUxDES}'s best fit from \cref{fig:dndz_bias}, as well as the fixed SKADS \citep{2008MNRAS.388.1335W}, an estimate based on the VLA-COSMOS \(\qty{3}{\giga\hertz}\) survey \citep{2017A&A...602A...5N,Smolcic2017}, and the redshift distribution of AGN from EMU Early Science data \citep{2022MNRAS.515.6046P} are shown as solid lines. The solid, salmon curve is estimated by cross-matching EMU \selavy\ islands with unselected \Euclid\ Q1 data as described in \cref{apdx:xmatch}. This latter estimate is likely biased towards the \Euclid\ redshift distribution as the cross-matching rate is below \(40\%\). The shaded grey area marks the redshift range not covered by \Euclid.}
\label{fig:Nz_fit}
\end{figure}

We use the \verb~zeus~ code \citep{2020arXiv200206212K,2021MNRAS.508.3589K} with six random walkers until the \citet{1992StaSc...7..457G} statistic attains \(1-\hat R < 5\text{\textperthousand}\), indicating good convergence and mixing. The resulting constraints on the EMU redshift distribution are presented in \cref{fig:Nz_triangles,fig:Nz_fit}, as well as in \cref{tab:Nz_maps}. We observe a lower MAP and posterior mean for \(z_0\) than \citet{Saraf:EMUxDES}, but consistent values for \(z_\mathrm{tail}\) and \(\gamma\), resulting in agreement within \(2\,\sigma\) in \(p_\mathrm{EMU}(z)\) over the entire redshift range. We observe very good (within \(1\,\sigma\)) agreement with the redshift distribution of AGN selected from the Million Quasars \citep[Milliquas;][]{2019arXiv191205614F,2019yCat.7283....0F} selected in the Large Magellanic Cloud (LMC) field for cross-matching with EMU Early Science data \citep[][cf.\ the brown curve in \cref{fig:Nz_fit}]{2022MNRAS.515.6046P}.
However, one limitation is that we assumed the bias model from \citet{Saraf:EMUxDES}; ultimately, we need to test \(p_\mathrm{EMU}(z)\) whilst jointly fitting for the galaxy bias.

With the physically motivated parameterisation, we observe two distinct peaks in \(p_\mathrm{EMU}(z)\), a feature that the other two parameterisations cannot capture. We find a MAP of \(q = 3.17_{-0.56}^{+2.65}\), in line with  
expectations from the \citet{AsoreyParkinson} study on the evolution of the proportion of SFGs and AGN with flux density cuts. We note a degeneracy between \(q\) and \(a\). Furthermore, the parameter \(a\) is constrained primarily by a prior that lacks a clear physical motivation and extending its prior range does not alleviate this prior dependence. Since \(a\) becomes influential at higher redshifts, we anticipate obtaining tighter constraints in our future re-analysis that will incorporate EMU auto-spectrum data, which extends beyond the \Euclid\ maximum redshift of \(z=2.5\). Overall, the primary peak of this physically motivated model is located at \(z=0.78\pm 0.11\), which is lower than the peak found by \citet{Saraf:EMUxDES} and in our baseline fit, namely \(z = 1.50 \pm 0.21\). Conversely, this peak is at a higher redshift than \(z=0.375\) predicted by \citet{2008MNRAS.388.1335W}. 

Despite these differences, the reconstructed \(p_\mathrm{EMU}(z)\) from the physically motivated model generally agrees within \(2\,\sigma\) with that obtained from the non-parametric cubic spline, particularly regarding the primary redshift peak and the overall shape at higher redshifts. The main discrepancy arises at low redshifts, where the physically motivated model can exhibit a secondary peak that is not evident in the smoother cubic spline reconstruction, which closely follows the baseline fit instead. 

Our reconstruction of \(p_\mathrm{EMU}(z)\) assuming the physically motivated redshift model is also consistent well with the redshift distributions of the Karl G.\ Jansky Very Large Array (VLA) cross-matched against the Cosmic Evolution Survey at \(\qty{3}{\giga\hertz}\) \citep[COSMOS \(\qty{3}{\giga\hertz}\);][]{2017A&A...602A...5N, 2017A&A...602A...6S}, which, in turn, agrees well with the \(p_\mathrm{EMU}(z)\) inferred from the MeerKAT International GHz Tiered Extragalactic Exploration (MIGHTEE) Survey, \citep{2024MNRAS.527.3231W}, once rescaled to match our frequency and flux density cut. We plot the COSMOS \(\qty{3}{\giga\hertz}\) redshift distribution for comparison in \cref{fig:Nz_fit}. The redshifts of MIGHTEE sources are based on visual cross-match and a likelihood ratio analysis applied to the UltraVISTA \(K_{\rm s}\) band selected catalogue \citep{2012A&A...544A.156M}. Their analysis reveals a significantly larger number of AGN at \(z \approx 1\) than predicted by simulations. The COSMOS \(\qty{3}{\giga\hertz}\) redshift distribution also roughly follows that of EMU Early Science data cross-matched with spectroscopic AGN observations by \citet[][cf.\ the violet curve in \cref{fig:Nz_fit}]{2022MNRAS.515.6046P}. Note that only their Gold sample down to \(\qty{500}{\micro\jansky}\) is complete, which explains why we tend to see more objects at higher redshifts in all parameterisations considered here.

Our analysis successfully constrains model parameters for the EMU source redshift distribution, demonstrating the power of cross-correlation techniques with \(\Euclid\) data. Whilst these model distributions provide a good fit to our observed cross-spectra, we must note that the resulting \(p_\mathrm{EMU}(z)\) is highly sensitive to the assumed bias model, \(b_\mathrm{EMU}(z)\), and \(n_\mathrm{EMU}(z)\) parameterisation due to the inherent \(p_\mathrm{EMU}(z)\,b_\mathrm{EMU}(z)\) degeneracy. This limitation is particularly relevant since we lack direct spectroscopic redshift data for these sources, and our different parameterisations yield \(p_\mathrm{EMU}(z)\) profiles with dramatically different peak locations (e.g., the \citealt{Saraf:EMUxDES} baseline peaks significantly higher than the physically motivated model). We are optimistic that future work incorporating EMU auto-spectrum data will be essential to help lift this degeneracy, allowing for a more robust, model-independent constraint on the true \(p_\mathrm{EMU}(z)\) and \({b_\mathrm{EMU}(z)}\) functions.

\section{Conclusions}\label{sec:conclusions}
In this pathfinder study, we have presented the first measurement of the harmonic-space cross-correlation power spectrum between radio-continuum sources from the EMU Main Survey and the galaxies in the Q1 Data Release from the \Euclid satellite mission. Utilising data from EDF-S, our primary objective was to detect and characterise this cross-correlation signal, assess its robustness against systematic effects, and compare our measurements with theoretical predictions.

We constructed galaxy number count maps from two independently generated EMU radio source catalogues (\selavy\ and \pybdsf) and two \Euclid photometric redshift samples (\pgal\ and \pstar). Employing the pseudo-\(C_\ell\) formalism with \texttt{NaMaster}, we measured the cross-spectra across a broad range of angular multipoles, \(\ell\in\left[2,801\right]\). Our analysis yields several key conclusions:
\begin{enumerate}
\item We report a high-significant detection of the cross-correlation signal, with a significance consistently exceeding \(8\,\sigma\) across all tested models and data sets (see \cref{tab:phz_chi2}). This robust detection confirms the strong synergy between deep radio continuum and optical/near-infrared galaxy surveys in tracing the LSS of the Universe.
\item The measured cross-spectra derived from the two independent EMU source catalogues (\selavy\ and \pybdsf) agree within \(1\,\sigma\) (see \cref{fig:cls_fidu}). This demonstrates that the detected cross-correlation is highly robust against the choice of source-finding algorithm, mitigating a key potential systematic uncertainty in radio surveys.
\item Our measured cross-correlation signal is consistent with theoretical models based on a fiducial \(\Lambda\)CDM cosmology and various redshift distribution and bias models for EMU sources, previously constrained by cross-correlations with other optical surveys and CMB lensing. We found an overall amplitude \(A_b\) consistent with unity across various models and \(\ell_\mathrm{max}\) ranges, validating the consistency of our measurements with established theoretical expectations for radio and optical galaxy populations. Furthermore, our derived values for the overall amplitude for the \Euclid samples are in excellent agreement with a sister analysis cross-correlating \Euclid data with CMB lensing (Fabbian et al., in prep.).
\item We assessed the contribution of cross-shot noise to the signal and found it to be subdominant (see \cref{ssec:shot_noise}). The fit for a separate shot-noise parameter was not well-constrained, confirming that common objects between the surveys do not significantly bias the clustering signal at the scales probed.
\item Moving beyond fixed theoretical assumptions, we used tomographic cross-correlations with three \(\Euclid\) redshift bins to constrain three different parameterisations (baseline, physically motivated, and cubic spline) for the EMU source redshift probability density function, \(p_\mathrm{EMU}(z)\), whilst fixing the EMU bias model to \(1.81/D(z)\). The resulting \(p_\mathrm{EMU}(z)\) profiles are highly model-dependent. For instance, the baseline fit results in a higher redshift peak compared to the physically motivated model, which aligns better with external data from COSMOS \(\qty{3}{\giga\hertz}\) and MIGHTEE. Crucially, the good fit obtained by all models, despite their different \(p_\mathrm{EMU}(z)\) shapes, highlights the strong degeneracy between \(p_\mathrm{EMU}(z)\) and \(b_\mathrm{EMU}(z)\). Only with additional information, such as the EMU auto-spectrum, can this degeneracy be lifted to determine accurately the true \(p_\mathrm{EMU}(z)\) and \(b_\mathrm{EMU}(z)\) of the radio source population.
\end{enumerate}

This work successfully establishes a statistically significant and robust cross-correlation between EMU and \Euclid data in a deep field. This serves as a crucial validation of the methodologies for combining these large-volume data sets and paves the way for future large-scale cosmological analyses. The ability to constrain the redshift distribution of radio sources through clustering redshifts is a powerful tool for unlocking the full cosmological potential of EMU.

Future work will leverage the full statistical power of the upcoming EMU Main Survey and the \Euclid Wide Field data releases to perform comprehensive tomographic analyses on an about 250 times larger overlap region, leading to an improvement of the cross-spectrum covariance of about a factor of \(\sim16\), and, hence, tighter constraints on cosmological parameters and a deeper understanding of galaxy evolution across cosmic time.
In addition, several current limitations could be addressed in future analyses. First, including the EMU and \Euclid auto-spectra will help disentangle galaxy distribution-bias degeneracies and improve the characterisation of the bias models for each survey. Furthermore, improving our understanding of the systematic effects in the data sets will allow us to extend cross-correlation measurements to smaller angular scales, providing a more direct probe of cross-shot noise that can be directly compared with results from the cross-match of sources. Finally, combining these data with CMB lensing cross-correlations of both \Euclid and EMU will offer a powerful, independent test of the growth of the LSS and will further tighten constraints on cosmological and astrophysical parameters. Together, these improvements will enable a more complete and self-consistent multi-tracer characterisation of the cosmic web.
%
\begin{acknowledgements}
We would like to thank David Alonso for his valuable comments and insightful discussions. GP, BB-K, and SC acknowledge support from the Italian Ministry of University and Research (\textsc{mur}), PRIN 2022 `EXSKALIBUR – Euclid-Cross-SKA: Likelihood Inference Building for Universe's Research', Grant No.\ 20222BBYB9, CUP D53D2300252 0006, and from the European Union---Next Generation EU.  BB-K furthermore acknowledges support from INAF for the project `Paving the way to radio cosmology in the SKA Observatory era: synergies between SKA pathfinders/precursors and the new generation of optical/near-infrared cosmological surveys', CUP C54I1900105 0001. JA acknowledges the support of the grant PGC2022-126078NB-C21 funded by MCIN/AEI/10.13039/ 50110001103 and Diputación General de Aragón-Fondo Social Europeo (DGA-FSE) grant 2023-E21-23R funded by Gobierno de Aragón. CLH acknowledges support from the Hintze Family Charitable Foundation through the Oxford Hintze Centre for Astrophysical Surveys. KT is supported by the STFC grant ST/W000903/1 and by the European Structural and Investment Fund. KT further acknowledges support by the European Union’s Horizon
Europe research and innovation programme under the Marie Sklodowska-Curie Postdoctoral Fellowship Programme, SMASH co-funded under the grant agreement No. 101081355. MR acknowledges support by the Italian Ministry of University and Research (\textsc{mur}) via the PRIN 2022 Project No.\ 20228WHTYC – CUP: D53C24003550006 and by the Research grant TAsP (Theoretical Astroparticle Physics) funded by \textsc{infn}.\\

This work made use of Pleiadi \citep{2020ASPC..527..307T,2020ASPC..527..303B}, a computing infrastructure installed and managed by INAF-USCVIII, under the proposal `Simulations and Analyses Across the Spectrum - Setting the Stage for Euclid-SKAO Synergies'. We also acknowledge the use of the computational resources of CESAR at BIFI Institute (University of Zaragoza). Results have been obtained and presented using the \texttt{HEALPix} \citep{2005ApJ...622..759G,Zonca2019}, \texttt{NaMaster} \citep{2019MNRAS.484.4127A}, \texttt{GLASS} \citep{Tessore_2023}, \texttt{pyccl} \citep{Chisari_2018}, \texttt{numpy} \citep{Harris_2020}, \texttt{matplotlib} (which provides \texttt{pyplot}) \citep{Hunter_2007}, \texttt{zeus} \citep{2021MNRAS.508.3589K}, and \texttt{getdist} \citep{2019arXiv191013970L} packages. We thank Steve Cunnington for sharing plotting scripts with us.\\

This scientific work uses data obtained from Inyarrimanha Ilgari Bundara, the CSIRO Murchison Radio-astronomy Observatory. We acknowledge the Wajarri Yamaji People as the Traditional Owners and native title holders of the Observatory site. CSIRO's ASKAP radio telescope is part of the Australia Telescope National Facility (https://ror.org/05qajvd42). Operation of ASKAP is funded by the Australian Government with support from the National Collaborative Research Infrastructure Strategy. ASKAP uses the resources of the Pawsey Supercomputing Research Centre. Establishment of ASKAP, Inyarrimanha Ilgari Bundara, the CSIRO Murchison Radio-astronomy Observatory and the Pawsey Supercomputing Research Centre are initiatives of the Australian Government, with support from the Government of Western Australia and the Science and Industry Endowment Fund.\\

This paper includes archived data obtained through the CSIRO ASKAP Science Data Archive, CASDA (\url{http://data.csiro.au}).\\

\AckEC  \\
\AckQone(\cite{Q1cite})

\end{acknowledgements}

%
%

\bibliography{Euclid, Q1, EMU_main, EMUxEuclid} 

%

\begin{appendix}
  \onecolumn 
\section{Covariance estimation}
\label{apdx:cov}
Given the small survey area covered by the two data sets, it is crucial to assess the stability of the covariance estimation, particularly with respect to the inversion of the mode-coupling matrix. Even though we adopted large band powers with a width of \(\Delta\ell = 100\), it is still important to verify that the pipeline does not introduce biases in the estimation of the power spectra or their covariance.

To this purpose, we generated \(N_{\rm sim} = 100\) simulations of \Euclid \pgal\ and EMU \pybdsf\ mock galaxy catalogues using the Generator for Large-Scale Structure \citep[\texttt{GLASS},\footnote{\url{https://glass.readthedocs.io/v2024.1/index.html}.}][]{Tessore_2023} assuming the same underlying cosmology as in \cite{2020A&A...641A...6P}.
For each simulated catalogue, we constructed the galaxy count maps, which are related to the overdensity contrast \(\Delta\) via \cref{eq:delta_g}, and computed the cross-spectra, \(\tilde{C}_\ell^\times\), using the same pipeline adopted in the main analysis.
The covariance obtained from these simulated cross angular power spectra is given by
\begin{equation}
\label{cov_sims}
    \Sigma_{\rm sim} = \frac{1}{N_{\rm sim}-1}\,\sum_{i=1}^{N_{\rm sim}}{\vec d}_i\,{\vec d}_i^\dagger\;,
\end{equation}
where \({\vec d}_i = {\vec s}_i - \langle{\vec s}\rangle\), and \({\vec s}_i\) are the measured \(\tilde{C}^\times_\ell\)'s from the \(i\)th simulated map, and \(\langle{\vec s} \rangle\) is the mean over all simulations. 

\begin{figure}[h]
\centering
\includegraphics[width=0.6\linewidth]{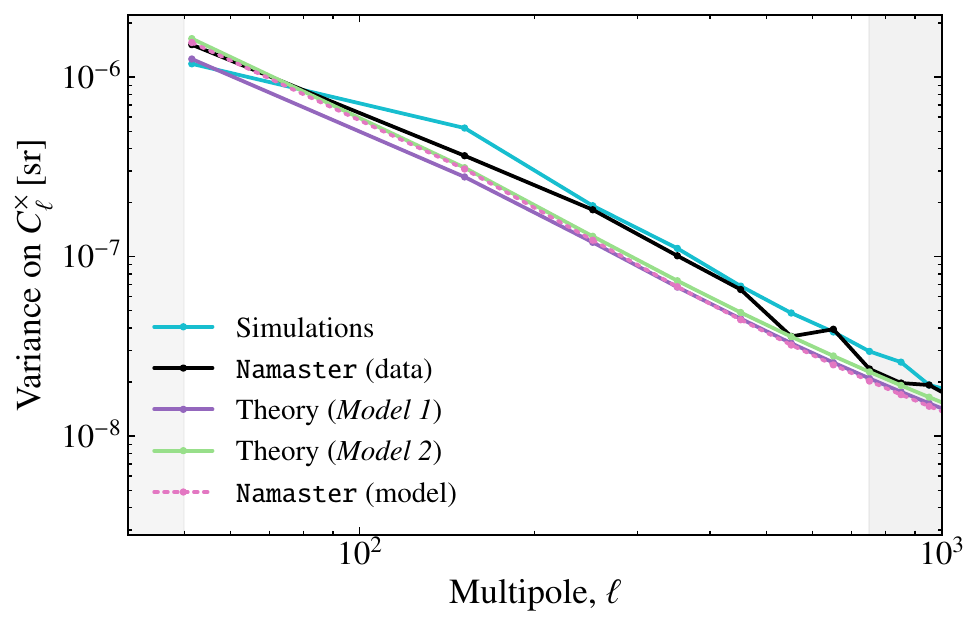}
\caption{Comparison of the estimated variance on the cross-spectrum \(\smash{C_\ell^\times}\). Different colours and markers indicate the covariance estimation method: simulations (cyan), analytic expressions (violet and green), Gaussian covariance from {\tt NaMaster} using theoretical \(\smash{C_\ell^\times}\)'s as input (dashed, pink), and Gaussian covariance from {\tt NaMaster} using measured angular power spectra (black). Grey shaded regions indicate angular scales excluded in our baseline configuration (cf.\ \cref{fig:cls_fidu}).}
\label{fig:sigma_sims}
\end{figure}

Then, we compared the covariance obtained from \cref{cov_sims} with different estimation of the covariance matrix: the one derived from the data and the theoretical predictions based on the model harmonic-space power spectra computed via \citep[see e.g.][]{Saraf_2021}
\begin{equation}
    {\rm Cov}\left(C_\ell^{AB},C_{\ell^{\prime}}^{CD}\right) = \frac{1}{(2\,\ell +1)\,\Delta\ell\,f_{\rm sky}^{AB}\,f_{\rm sky}^{CD}}\,\left[ f_{\rm sky}^{AC,BD}\,\sqrt{C_\ell^{AC}\,C_{\ell^{\prime}}^{AC}\,C_\ell^{BD}\,C_{\ell^{\prime}}^{BD}} +
    f_{\rm sky}^{AD, BC}\,\sqrt{C_\ell^{AD}\,C_{\ell^{\prime}}^{AD}\,C_\ell^{BC}\,C_{\ell^{\prime}}^{BC}}
    \right]\,\delta_{\ell\ell^{\prime}}\;,\label{eq:cov}
\end{equation}
where \( \{A, B, C, D\} \in \{\Euclid, \text{EMU}\} \), and \( \smash{f_{\rm sky}^{AB}} \) denotes the fraction of sky commonly observed by surveys \( A \) and \( B \). The term \( \smash{f_{\rm sky}^{AC,BD} }\) is defined as the geometric mean of the corresponding overlapping sky fractions, viz.\ \(\smash{f_{\rm sky}^{AC,BD} = (f_{\rm sky}^{AC}\,f_{\rm sky}^{BD})^{1/2}}\). The results are summarised in \cref{fig:sigma_sims}, where the ordinates show the square root of the diagonal elements of the \(\ell\)-\(\ell'\) covariance matrix of \cref{eq:cov}.

We find that the covariance estimates derived from simulations and from the data are in good mutual agreement, further supporting the robustness of our  pipeline, especially in the regime probed by this analysis. Nonetheless, both simulations and data covariances are consistently larger than the values predicted by the theoretical model. This discrepancy may affect some of the results discussed in \cref{sec:results}.
Specifically, we verified that using the analytical covariance would increase the detection significance by approximately \(12\%\) (as computed via Eq.\ \ref{eq:significance}, see \cref{tab:phz_chi2}), whilst having negligible impact on the joint constraints on \(\{A_b, C_{\rm shot}^\times\}\).

Given that the analytical covariance might underestimate the impact of residual systematics in these first \Euclid and EMU data sets -- and considering that our main results are not significantly affected by using the data covariance -- we make the conservative choice of adopting the latter in our main analysis. We leave a more detailed investigation of the covariance matrix to future works. 

\section{Positional cross-matching}
\label{apdx:xmatch}
\begin{figure}
\centering
\includegraphics[width = 0.65\columnwidth]{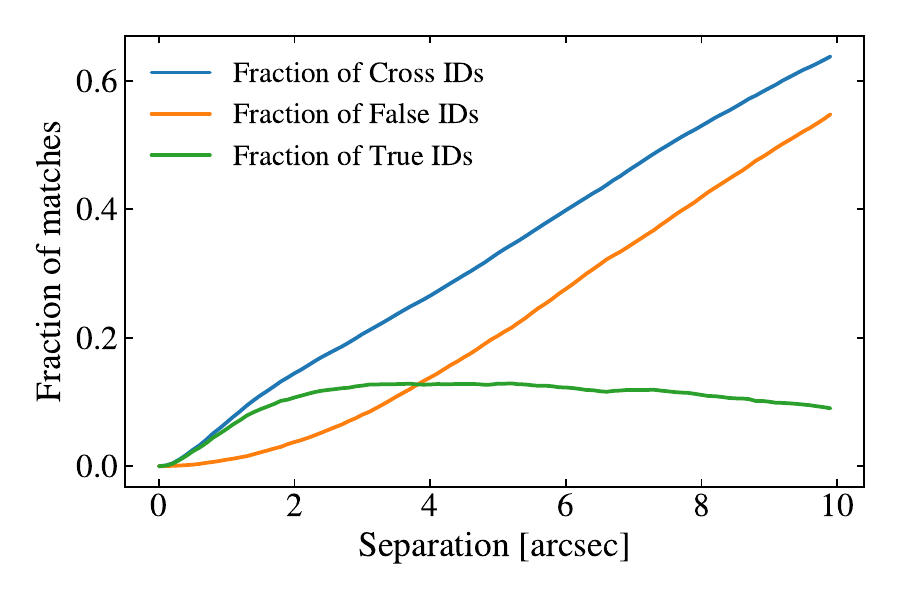}
\caption{Fractional cross-matching of \pybdsf\ and \pgal\ sources. The blue curve shows the total fraction of \pybdsf\ and \pgal\ objects cross-identified as a function of the maximum angular separation. The orange curve provides an estimate of the false cross-identification rate, determined by shifting the \pybdsf\ catalogue positions by \(\ang{;1}\) and repeating the match. The final estimated true cross-match fraction (green) is calculated as the difference between the total fraction and the false match fraction.}
\label{fig:xmatch}
\end{figure}

The obvious choice for estimating the cross-shot noise and EMU redshift distribution is to cross-identify objects that are common to both EMU and \Euclid, henceforth dubbed `cross-IDs'. However, this task is not as straightforward as it seems, and there are currently several EMU projects working to cross-identify EMU sources with various other surveys. We present here a simple cross-identification algorithm based solely on the positions of objects in either catalogue. Using standard \texttt{astropy}\footnote{\texttt{\url{https://www.astropy.org}}.} functions, we find for each EMU source the closest \Euclid galaxy, and measure their angular separation. If the angular separation is below a threshold (for instance, \(\ang{;;3}\)), we consider this pair to be the same object. Due to the angular extent of the sources, to pointing, and to astrometric errors, choosing a threshold that is too small will miss true cross-identifications. On the other hand, a too large threshold will cross-match objects that are just by chance close in angular position but actually unrelated and potentially far apart along the observer's line of sight. A simple way to estimate these false cross-IDs was also used by \citet{2021PASA...38...46N}, and consists of shifting the positions of one survey by \(\ang{;1}\) and repeating the cross-matching process. All cross-IDs found with the shifted catalogue are false as long as the separation threshold is significantly smaller than the shift. 

We show in \cref{fig:xmatch} the fraction of cross-IDs between the true \pybdsf\ and \pgal\ catalogues (blue curve), along with the cross-IDs found after shifting the \pybdsf\ positions (orange curve), as a function of the maximum allowed angular separation. We estimate the fraction of true cross-IDs as the difference between the two quantities (green curve). We see this difference plateauing, with a small decrease, above \(\ang{;;3}\), indicating that we cannot safely cross-match more than \(12.9\%\) of \pybdsf\ sources with a \pgal\ counterpart. We see a similar behaviour when repeating this analysis with the \selavy\ catalogue, for which we can find cross-identify \(13.4\%\) of its objects in the \pgal\ sample. For the \pstar\ sample, containing about twice as many objects as \pgal, we can match up to \(29.6\%\) of \pybdsf\ and \(29.9\%\) of \selavy\ sources. The significantly lower cross-ID rate of \pgal\ with respect to \pstar\ can be explained by the fact that the former selection criterion discards AGN.

These numbers seem low compared to cross-ID rates found by \citet{2021PASA...38...46N}. This can be explained by two main effects. Firstly, \citet{2021PASA...38...46N} restrict the analysis to simple one-component sources, which are about \(80\%\) of all sources, and which are usually within \(\ang{;;1}\) of the host. In the sample used throughout this work, including multi-component islands, these islands can be larger, and will also include confused sources, etc, and using a \(\ang{;;3}\) search radius from the centroid of the island may be too optimistic. However, \Euclid's high survey density prevents increasing the search radius without being affected by a large number of false cross-IDs. Secondly, the EMU Pilot Survey employed in \citet{2021PASA...38...46N}  has better positional accuracy than the EMU Main Survey. Efforts are underway to reprocess the main survey fields to fix this.

In any case, our cross-match numbers are too low to obtain a reliable estimate of EMU's redshift distribution. This inadequacy provides the core rationale for the approach adopted in the main body of this work, where we use the cross-spectrum with the reliable \Euclid redshift distribution as a probe to statistically constrain the redshift distribution of the full EMU sample (clustering redshifts).
For the sake of comparison, we download a catalogue containing all objects observed by \Euclid in EDF-S without any selection applied. Due to the large data volume, we restrict this analysis to a \(4\,\deg^2\) patch (14\% of the EDF-S footprint). Given the even higher source density of this all-encompassing catalogue, we restrict the angular separation threshold to \(\ang{;;1.7}\), corresponding to the peak of the difference between total and false cross-IDs.

Out of the 3412 EMU islands in that \(4\,\deg^2\) patch, we can cross-match 2116 \Euclid\ objects, of which we expect 853 to be falsely identified based on the shifting method described above, corresponding to a true cross-ID rate of \(37\%\). We show the redshift distribution of the cross-matched objects in \cref{fig:Nz_fit}. We notice some resemblance with the redshift distribution inferred from the SKADS simulation \citep{2008MNRAS.388.1335W}, albeit with a more pronounced peak at \(z = 0.24\) and a suppressed high-redshift tail. Since we cross-match with all observed objects, we can also look at the distribution of object types: \(70.0\%\) of cross-IDs are identified as galaxies, which is less than for the total \Euclid\ sample with \(81.9\%\) galaxies; contrarily, the cross-matched sample shows a significantly higher fraction of quasars (14.9\%) than the total sample (2.3\%). This can be expected since the \Euclid\ Q1 classifier classes luminous AGN as quasars \citep{Q1-SP026}. The secondary peak of the redshift distribution in \cref{fig:Nz_fit} at \(z=1.44\) corresponds to the peak of the redshift distribution of EMU islands matched with \Euclid\ QSO/AGN. Interestingly, \(4\) EMU sources are identified as stars, i.e., \(0.19\%\) of cross-IDs. This is significantly less than the \(2.3\%\) rate of stars in the total \Euclid\ sample and in line with the fraction of stars in LoTSS \citep{2023A&A...670A.124C}, underlining our statement in \cref{fn:stars}.

To conclude, we reiterate that, whilst this positional cross-match exercise provides valuable insight into object classification and expected source populations, the fundamental limitation remains the low cross-ID rate. The redshift distribution derived from this small, selected fraction is unequivocally biased, prioritising sources detectable by \Euclid and failing to represent the true statistical distribution of the full EMU source catalogue. This confirms the necessity of adopting the more robust clustering redshift approach for our main analysis.

\section{Counts-in-cell statistics}
\label{sec:cic}

In \cref{ssec:shot_noise}, we introduced the clustering parameter \(n_\mathrm{c}\) as a diagnostic for the Poissonianity of our galaxy samples. Whilst \(n_\mathrm{c}\) provides a single summary statistic, a more detailed understanding of the source count distribution can be gained through counts-in-cell (CIC) statistics. This method involves counting the number of sources within predefined spatial cells (in our case, \texttt{HEALPix} pixels) and then analysing the resulting distribution of counts.
For a truly random, uncorrelated distribution of sources, the counts in cells are expected to follow a Poisson distribution. However, as discussed in \cref{ssec:shot_noise}, high-resolution radio surveys can exhibit overdispersion due to the fragmentation of single physical sources into multiple detected components, leading to non-Poissonian behaviour. In such cases, the negative binomial distribution often provides a more accurate statistical model for the observed counts.

\begin{figure}[h]
\centering
\includegraphics[width = 0.65\columnwidth]{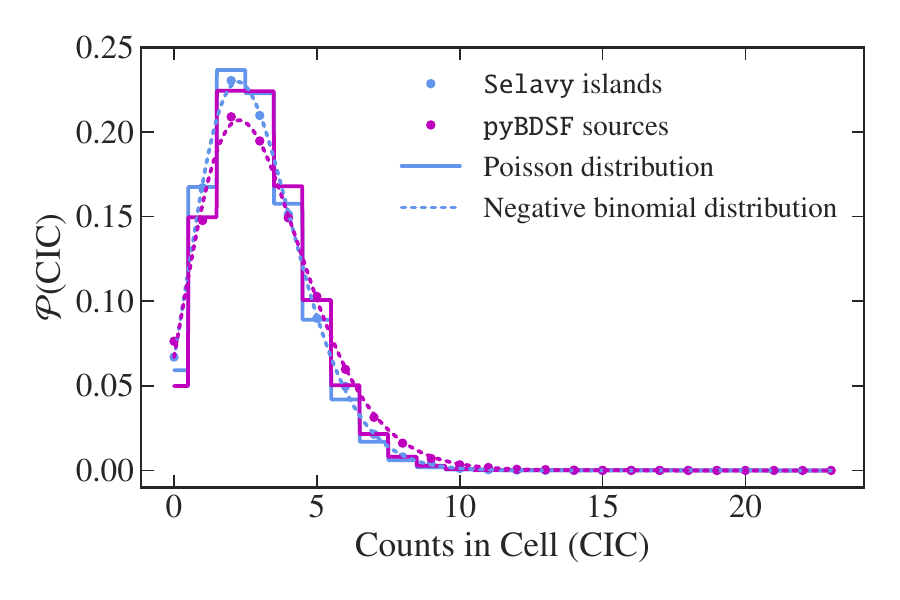}
\caption{Counts of \selavy\ islands (blue) and \pybdsf\ sources (purple) in \texttt{HEALPix} cells with \(N_{\rm side}= 1024\). The histograms show the Poissonian distribution fit to the data points in the same colour, whereas the dashed curves indicate a fit to a negative binomial distribution.}
\label{fig:cic}
\end{figure}

\Cref{fig:cic} presents the normalised frequencies of CIC for both the \selavy\ island sample (blue markers) and the \pybdsf\ source catalogue (purple markers) within \texttt{HEALPix} cells at \(N_{\rm side}=1024\). Overlaid on these histograms are the best-fit Poisson distributions (solid histogram) and negative binomial distributions (dashed curves). As anticipated from our \(n_\mathrm{c}\) values in \cref{ssec:shot_noise}, the \selavy\ island counts closely follow a Poisson distribution (\(n_\mathrm{c} = 1.06\)), with the solid blue curve providing an excellent fit to the data points. On the other hand, for the \pybdsf\ source catalogue, whilst still reasonably approximated by a Poisson distribution, the counts exhibit a slightly broader spread (\(n_\mathrm{c} = 1.28\)). This mild overdispersion is better captured by the negative binomial distribution, which provides a visually superior fit to the \pybdsf\ data, particularly in the tail of the distribution.

This detailed CIC analysis confirms our earlier assessment that the \selavy\ sample is nearly Poissonian, whilst the \pybdsf\ sample shows a slight deviation. Despite this, for the angular scales and objectives of the main analysis, the Poissonian approximation for shot noise in both catalogues is considered sufficient. However, for future work, especially when extending to smaller scales or performing auto-spectrum analyses where these effects become more pronounced, a more rigorous treatment of non-Poissonian shot noise, potentially using the negative binomial model, will be crucial.

\section{Scale cuts and  nonlinearities}\label{sec:lmax_nonlin}
We have argued in \cref{ssec:theory_met.theory} that a linear bias model is sufficient for our analysis, as we consider only quasi-linear scales. Following the approaches of \citet{Efstathioukmax}, \citet{SMith2003}, and \citet{Blakekmax}, we define the maximum wavenumber \(k_\mathrm{max}^\mathrm{NL}\) using the model power spectrum to calculate the dimensionless variance of mass fluctuations. This is given by
\begin{equation}
    \sigma^2(x, z) = \frac{1}{2\,\pi^2}\,\int \de k\, k^2\,P(k, z)\,\left\vert W(k\,x)\right\vert^2\;,
\end{equation}
where \(W\) is the filter function chosen to impose a cut-off at \(k_\mathrm{max}^\mathrm{NL}\). This cut-off corresponds to the comoving scale \(R\) at which fluctuations become  nonlinear, with the filter matching a half-wavelength to the diameter of the spheres. We thus define \(k_\mathrm{max}^\mathrm{NL}\) by solving
\begin{equation}
    \sigma^2(k_\mathrm{max}^\mathrm{NL}, z) = \frac{1}{2\,\pi^2}\,\int_0^{k_{\max}^\mathrm{NL}}\de k\,k^2\,P(k, z)\;.
\end{equation}
Whilst \citet{Blakekmax} and \citet{Seokmax} choose to obtain \(k_\mathrm{max}^\mathrm{NL}\) by solving \(\sigma(k_\mathrm{max}^\mathrm{NL}, z) = 1/2\), we adopt a more conservative approach, solving \(\sigma(k_\mathrm{max}^\mathrm{NL}, z) = \sigma_8/2\). This yields \(k_\mathrm{max}^\mathrm{NL}\) values comparable to those found by \citet{Seokmax}, where \(\sigma_8 = 1\) was assumed.

\begin{figure}
\centering
\includegraphics[width = \linewidth]{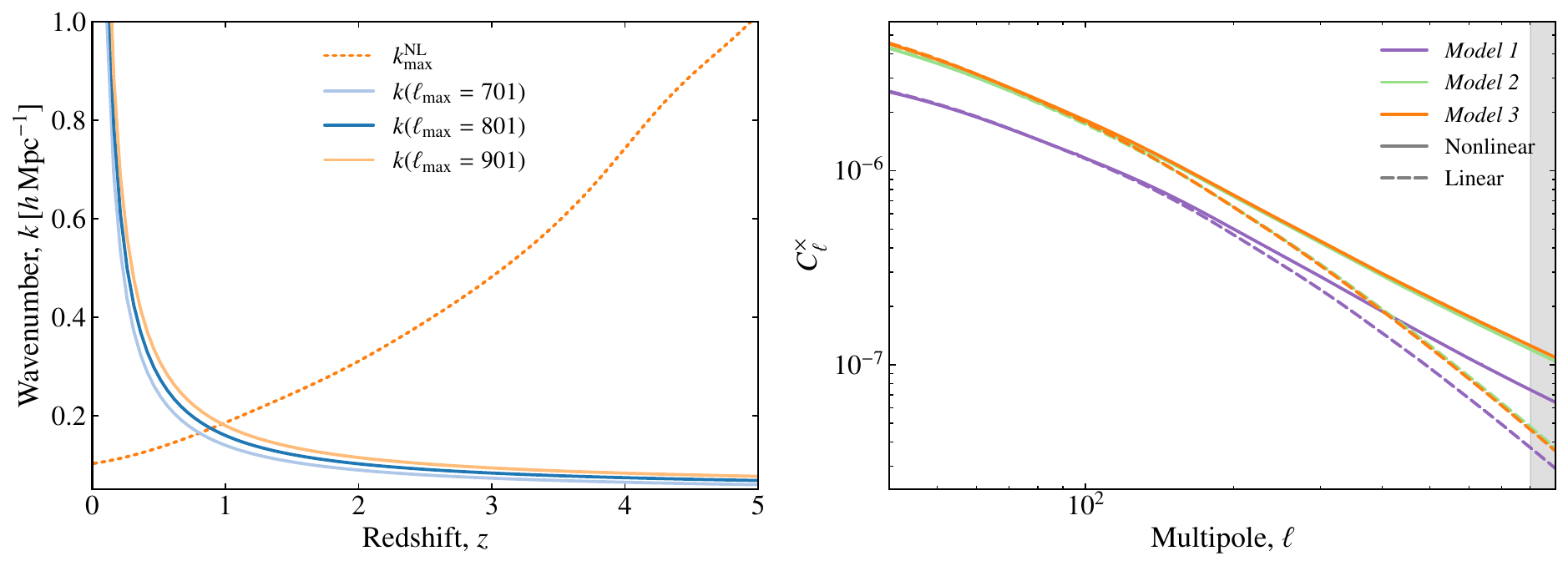}
\caption{{\it Left}: Wavenumbers corresponding to a multipole in the Limber approximation (cf.\ Eq.\ \ref{eq:Cl_Limber}), compared to the maximum wavenumber at which  nonlinear clustering becomes important. {\it Right}: Comparison of the nonlinear (solid) and linear (dashed) model power spectra for {\it Models 1}, \textit{2}, and \textit{3}, respectively, as defined in \cref{ssec:theory_met.theory}. Note that we summarise results only for \pgal\ given that \pstar\ shows a similar behaviour.}
\label{fig:kmax_limber}
\end{figure}

We plot \(k_\mathrm{max}^\mathrm{NL}\) in \cref{fig:kmax_limber} along with the wavenumbers 
\begin{equation}
    k(\ell) = \frac{\ell + 1/2}{r(z)}\;,
\end{equation}
corresponding to the \(\ell\) cuts considered in this work according to the Limber approximation (Eq.\ \ref{eq:Cl_Limber}). At low redshifts, even wide angles (or small \(\ell\)'s) correspond to small scales, i.e., large \(k\)'s. Thus, as \(z\to0\), any \(\ell\) falls within the nonlinear regime, as \(k(\ell)\) diverges. Furthermore, \(k_\mathrm{max}^\mathrm{NL}\) decreases as the Universe evolves, with clustering becoming  nonlinear at progressively larger scales. However, we also anticipate fewer objects at very low redshift (cf.\ \cref{fig:dndz_bias}). 

To assess whether our scale cuts ensure linearity, we therefore compute the effective redshift as
\begin{equation}
    z_\mathrm{eff} = \int \de z \, p(z)\, z\;.
\end{equation}
For our baseline \textit{Model 1}, as defined in \cref{ssec:theory_met.theory}, the mean EMU redshift is \(z_\mathrm{eff, EMU}=1.43\). The effective maximum wavenumber at this redshift amounts to \(k_\mathrm{eff, EMU} = \qty{0.12}{\hubble\per\mega\parsec}\),
which is a factor of two smaller than the wavenumber \(k_\mathrm{max,EMU}^\mathrm{NL} \simeq \qty{0.24}{\hubble\per\mega\parsec}\) where nonlinear effects become significant. We perform analogous calculations for \Euclid\ \pstar, finding \(z_\mathrm{eff,Euc}=0.87\) and, correspondingly, \(k_\mathrm{eff,Euc}= \qty{0.17}{\hubble\per\mega\parsec}\equiv k_\mathrm{max,Euc}^\mathrm{NL}\), i.e., it is precisely at the maximum linear wavenumber for \Euclid.
It is important to note that we do not consider the EMU and \Euclid\ auto-spectra in this work. The geometric mean of the effective redshifts for the cross-correlation is \(z_\mathrm{eff,\times}= 1.12\). At this redshift, we have \(k_\mathrm{eff,\times}=\qty{0.15}{\hubble\per\mega\parsec}<k_\mathrm{max,\times}^\mathrm{NL} =\qty{0.20}{\hubble\per\mega\parsec}\).
However, the use of a single effective redshift, whilst informative, can mask significant contributions from low-redshift regimes. To address this, we split the \Euclid\ sample into quartiles and compute the effective redshift of the lowest quartile, \(z_\mathrm{eff, q_1} = 0.25\). Here, our maximum multipole corresponds to \(k_\mathrm{eff, q_1} = \qty{0.52}{\hubble\per\mega\parsec}\), which is far into the nonlinear regime, delimited by \(k_\mathrm{max,q_1}^\mathrm{NL} =\qty{0.12}{\hubble\per\mega\parsec}\). Therefore, we also compare the linear and nonlinear model power spectra to check at what angular scale the two start to differ and plot them in the right-hand panel of \cref{fig:kmax_limber}. We observe significant nonlinear contributions at fairly low \(\ell \sim 300\). However, our theoretical predictions rely on the \texttt{halofit} prescription, which is specifically calibrated to model the matter power spectrum in the quasi-linear and nonlinear regimes, typically up to \(k \sim\qty{0.5}{\per\mega\parsec}\) in \(\Lambda\)CDM cosmology. The scale range \(\ell \in [2, 801]\) is therefore well within the validated range for \texttt{halofit} used in similar large-scale structure studies. Given that our fiducial models fit the measured data points well across the entire \(\ell\) range, we consider our current approach sufficient for the primary goal of establishing a robust cross-correlation signal and constraining our \(n(z)\,b(z)\) models. In future, more advanced cosmological analyses, we will more carefully evaluate the breakdown of \texttt{halofit} at the lowest redshifts and smallest scales, potentially necessitating alternative nonlinear extensions or incorporating more complex effects like baryon feedback, or stronger \(\ell\) or \(z\) cuts.

\section{Triangle plots}
\begin{figure}[h]
\centering
\includegraphics[width = 0.47\columnwidth]{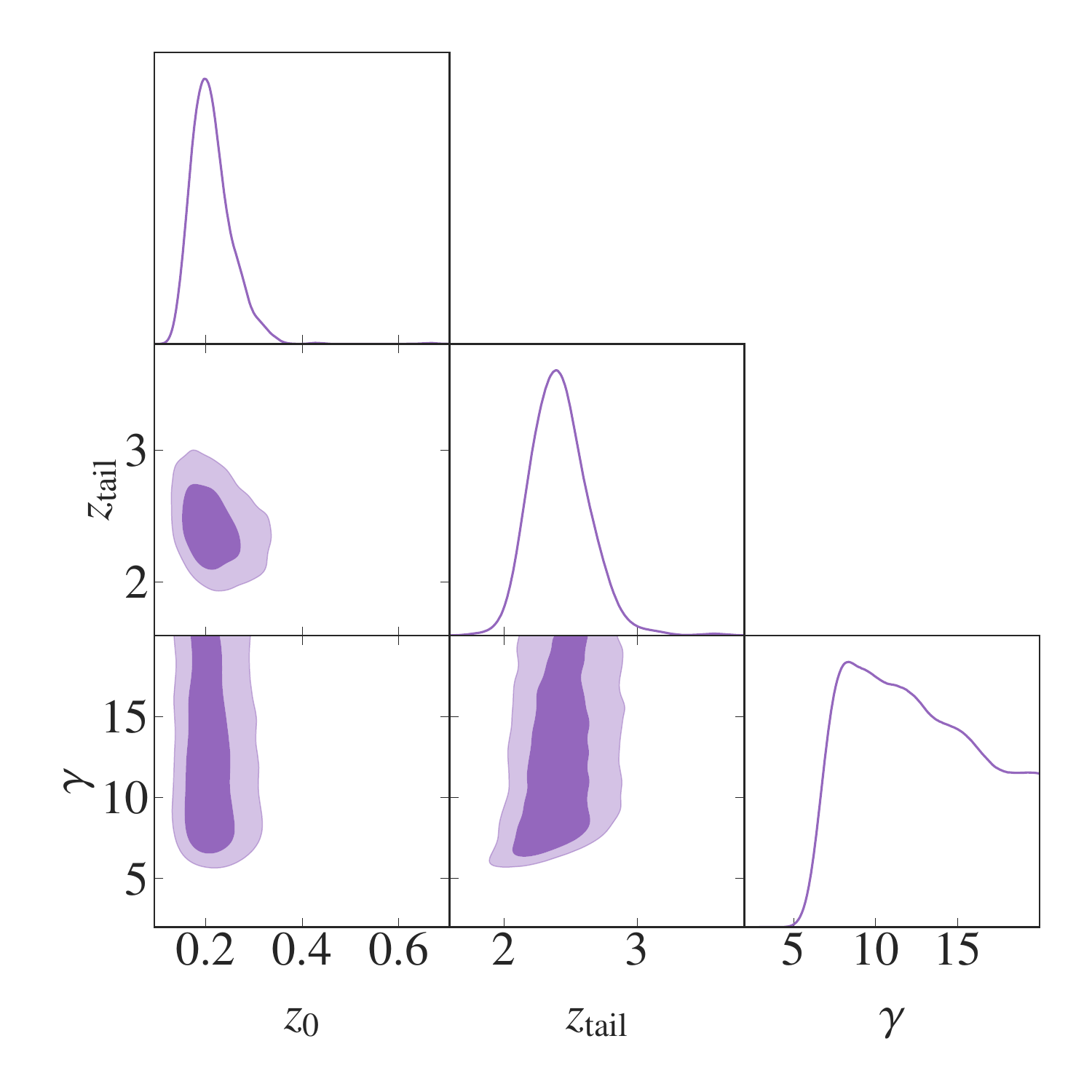}
\includegraphics[width = 0.47\columnwidth]{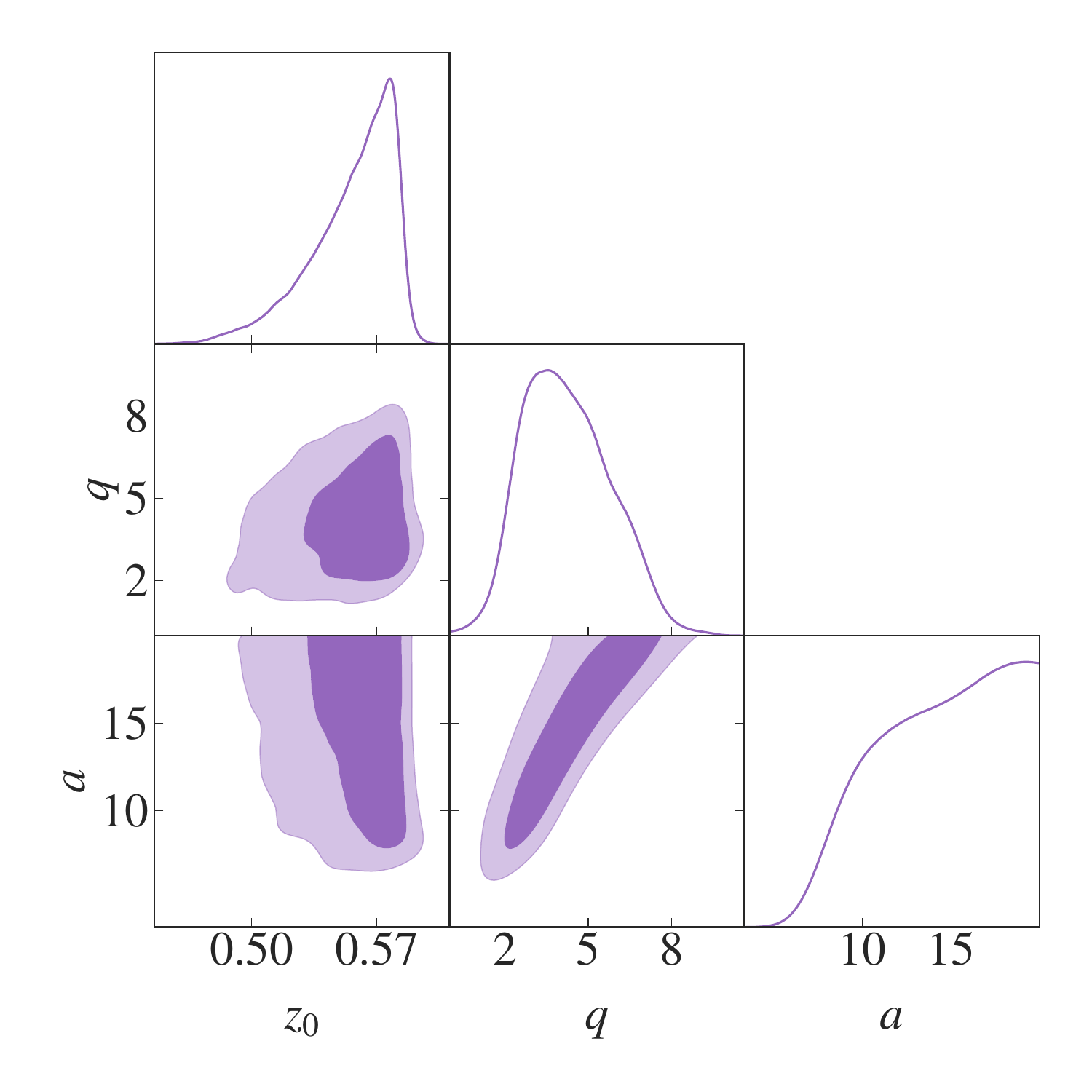}
\caption{Posterior distribution from fitting the fiducial \citep[][\textit{left}]{2021MNRAS.502..876A} and physically motivated \citep[][\textit{right}]{2024A&A...681A.105N} redshift distribution models to the EMU-\Euclid cross-spectra in three redshift bins.
}
\label{fig:Nz_triangles}
\end{figure}

This appendix presents the full posterior distributions of the parameters for the baseline and physically-motivated EMU source redshift distribution models discussed in \cref{sec:Nzfit}. The left-hand side of \cref{fig:Nz_triangles} displays the corner plot for the parameters of the baseline parameterisation \((z_0, z_\mathrm{tail}, \gamma)\) as defined in \cref{eq:model_nz_lotssdr1}. The marginalised 1D and 2D posterior distributions illustrate the correlations and uncertainties in these parameters. The best-fit values and their credible intervals are summarised in \cref{tab:Nz_maps}.
We notice, that our \(z_0\) posterior peaks significantly lower than that of \citet{Saraf:EMUxDES}. This corresponds to the more than \(2\,\sigma\) discrepancy in \(p_\mathrm{EMU}(z)\) at low redshifts shown in \cref{fig:Nz_fit}. As \(z_0\) and \(z_\mathrm{tail}\) have some level of degeneracy, our \(z_\mathrm{tail}\) is more than \(1\,\sigma\) lower than that of \citet{Saraf:EMUxDES}, even though our \(p_\mathrm{EMU}(z)\) results are largely consistent above redshifts of about 0.5. 

Similarly, the right-hand side of \cref{fig:Nz_triangles} shows the posterior distributions for the parameters of the physically motivated model \((z_0, q, a)\). These plots provide a detailed view of the parameter degeneracies, particularly between \(q\) and \(a\), and inform our discussion on the interpretation of this model's redshift distribution. The summarised statistics for these parameters can be found in \cref{tab:Nz_maps}.

\end{appendix}

\label{LastPage}
\end{document}